\documentclass[twocolumn]{aastex7}

\def\OI{O\,{\sc i}} 
\def\CaII{Ca\,{\sc ii}}

\usepackage[utf8]{inputenc}
\usepackage{textcomp}
\usepackage{caption}

\newcommand{\Mdot}[1]{$M_{\odot}$}
\newcommand{\Rdot}[1]{$R_{\odot}$}

\usepackage{CJK}
\usepackage{amssymb}
\begin{document}

\title{The Double-Peaked Calcium-Strong SN 2025coe: Progenitor Constraints from Early Interaction and Ejecta Asymmetries
}

\correspondingauthor{Aravind Pazhayath Ravi}
\email{apazhayathravi@ucdavis.edu}

\newcommand{\UA}{\affiliation{Steward Observatory, University of Arizona, 933 North Cherry Avenue, Tucson, AZ 85721-0065, USA}}
\newcommand{\GeminiNorth}{\affiliation{Gemini Observatory, 670 North A`ohoku Place, Hilo, HI 96720-2700, USA}}
\newcommand{\Catalyst}{\altaffiliation{LSSTC Catalyst Fellow}}
\newcommand{\Hubble}{\altaffiliation{NASA Hubble Fellow}}
\newcommand{\Monash}{\affiliation{School of Physics and Astronomy, Monash University, Clayton, Australia}}
\newcommand{\OzGrav}{\affiliation{OzGrav: The ARC Center of Excellence for Gravitational Wave Discovery, Australia}}
\newcommand{\UCSD}{\affiliation{Department of Astronomy \& Astrophysics, University of California, San Diego, 9500 Gilman Drive, MC 0424, La Jolla, CA 92093-0424, USA}}
\newcommand{\UCD}{\affiliation{Department of Physics and Astronomy, University of California, Davis, 1 Shields Avenue, Davis, CA 95616-5270, USA}}
\newcommand{\LCO}{\affiliation{Las Cumbres Observatory, 6740 Cortona Drive, Suite 102, Goleta, CA 93117-5575, USA}}
\newcommand{\UCSB}{\affiliation{Department of Physics, University of California, Santa Barbara, CA 93106-9530, USA}}
\newcommand{\Keck}{\affiliation{W.~M.~Keck Observatory, 65-1120 M\=amalahoa Highway, Kamuela, HI 96743-8431, USA}}
\newcommand{\CfA}{\affiliation{Center for Astrophysics \textbar{} Harvard \& Smithsonian, 60 Garden Street, Cambridge, MA 02138-1516, USA}}
\newcommand{\UNC}{\affiliation{Department of Physics and Astronomy, University of North Carolina, 120 East Cameron Avenue, Chapel Hill, NC 27599, USA}}
\newcommand{\CIERA}{\affiliation{Center for Interdisciplinary Exploration and Research in Astrophysics (CIERA), 1800 Sherman Ave., Evanston, IL 60201, USA}}
\newcommand{\NU}{\affiliation{Department of Physics and Astronomy, Northwestern University, 2145 Sheridan Road, Evanston, IL 60208, USA}}
\newcommand{\USask}{\affiliation{Department of Physics and Engineering Physics, University of Saskatchewan, 116 Science Place, Saskatoon, SK S7N 5E2, Canada}}
\newcommand{\UvA}{\affiliation{Department of Astronomy, University of Virginia, 530 McCormick Rd, Charlottesville, VA 22904, USA}}
\newcommand{\SU}{\affiliation{The Oskar Klein Centre, Department of Astronomy, Stockholm University, AlbaNova, SE-10691 Stockholm, Sweden}}
\newcommand{\Rutgers}{\affiliation{Department of Physics and Astronomy, Rutgers, The State University of New Jersey, 136 Frelinghuysen Rd, Piscataway, NJ 08854-8019, USA}}
\newcommand{\Tsinghua}{\affiliation{Physics Department, Tsinghua University, Beijing 100084, China}}
\newcommand{\Berkeley}{\affiliation{Department of Astronomy, University of California, Berkeley, CA 94720-3411, USA}}
\newcommand{\UCSC}{\affiliation{Department of Astronomy and Astrophysics, University of California, Santa Cruz, CA 95064, USA}}
\newcommand{\UF}{\affiliation{Department of Astronomy, University of Florida, 211 Bryant Space Science Center, Gainesville, FL 32611-2055, USA}}
\newcommand{\KITP}{\affiliation{Kavli Institute for Theoretical Physics, University of California, Santa Barbara, CA 93106, USA}}
\newcommand{\STScI}{\affiliation{Space Telescope Science Institute, 3700 San Martin Drive, Baltimore, MD 21218, USA}}
\newcommand{\Xinjiang}{\affiliation{Xinjiang Astronomical Observatory, Chinese Academy of Sciences, Urumqi, Xinjiang, 830011, China}}
\newcommand{\NRAO}{\affiliation{National Radio Astronomy Observatory, 520.0Edgemont Rd, Charlottesville VA 22903, USA}}

\begin{CJK*}{UTF8}{gbsn}

\author[0000-0002-7352-7845]{Aravind P. Ravi} \UCD \email{apazhayathravi@ucdavis.edu}

\author[0000-0001-8367-7591]{Sahana Kumar} \UvA \email{bsw2dc@virginia.edu}

\author[0009-0004-7268-7283]{Raphael Baer-Way} \UvA \NRAO
\email{bek5cw@virginia.edu}

\author[0000-0001-8818-0795]{Stefano Valenti} \UCD \email{valenti@ucdavis.edu}

\author[0000-0001-7132-0333]{Maryam Modjaz} \UvA
\email{vru7qe@virginia.edu}

\author[0009-0001-3767-942X]{Bart F. A. van Baal} \SU \email{barteld.vbaal@astro.su.se}

\author[0000-0001-8005-4030]{Anders Jerkstrand} \SU \email{anders.jerkstrand@astro.su.se}

\author[0000-0002-7937-6371]{Yize Dong (董一泽)} 
\CfA 
\email{yize.dong@cfa.harvard.edu} %

\author[orcid=0000-0003-3108-1328, gname=Lindsey, sname=Kwok]{Lindsey A. Kwok} \Hubble
\CIERA  \email{lindsey.kwok@northwestern.edu}

\author[orcid=0000-0002-0744-0047, gname=Jeniveve, sname=Pearson]{Jeniveve Pearson}
\UA \email{jenivevepearson@arizona.edu}

\author[orcid=0000-0003-4102-380X, gname=David, sname= Sand]{David J. Sand}
\UA \email{dsand@arizona.edu}

\author[orcid=0000-0002-1125-9187]{Daichi Hiramatsu} \UF \email{}

\author[0000-0003-3460-0103]{Alexei V. Filippenko}
\Berkeley \email{}

\author[orcid=0000-0003-0123-0062, gname=Jennifer, sname=Andrews]{Jennifer Andrews}
\GeminiNorth \email{Jennifer.Andrews@noirlab.edu}

\author[0000-0002-1895-6639]{Moira Andrews}
\LCO \UCSB \email{}

\author[orcid=0000-0002-6688-3307]{Prasiddha Arunachalam} 
\UCSC \email{}

\author[orcid=0000-0002-4924-444X, gname= Azalee, sname=Bostroem]{K. Azalee Bostroem} \Catalyst 
\UA \email{bostroem@arizona.edu}

\author[0000-0001-5955-2502]{Thomas G. Brink} 
\Berkeley \email{}


\author[orcid=0000-0002-9946-1477]{Liyang Chen}
\Tsinghua \email{chenly23@mails.tsinghua.edu.cn}

\author[orcid=0000-0003-0528-202X, gname=Collin, sname=Christy]{Collin Christy}
\UA \email{collinchristy@arizona.edu}

\author[orcid=0000-0002-5680-4660]{Kyle W. Davis} 
\UCSC \email{}

\author[orcid=0000-0003-1845-4900]{Ali Esamdin}
\Xinjiang \email{aliyi@xao.ac.cn}

\author[0000-0003-4914-5625]{Joseph Farah}
\LCO \UCSB \email{}

\author[orcid=0000-0002-2445-5275]{Ryan J. Foley} 
\UCSC \email{}

\author[orcid=0000-0003-2744-4755, gname=Emily, sname=Hoang]{Emily Hoang}
\UCD \email{emthoang@ucdavis.edu}

\author[orcid=0000-0002-0832-2974, gname=Griffin, sname=Hosseinzadeh]{Griffin Hosseinzadeh}
\UCSD \email{ghosseinzadeh@ucsd.edu}

\author[0000-0003-4253-656X]{D.\ Andrew Howell}
\LCO \UCSB \email{}

\author[orcid=0000-0002-9454-1742, gname=Brian, sname=Hsu]{Brian Hsu}
\UA \email{bhsu@arizona.edu}

\author[orcid=0009-0002-3079-1133]{Ruifeng Huang}
\Tsinghua \email{huangrf24@mails.tsinghua.edu.cn}

\author[orcid=0009-0003-9229-9942]{Abdusamatjan Iskandar}
\Xinjiang \email{abudu@xao.ac.cn}

\author[orcid=0000-0003-0549-3281, gname=Daryl, sname=Janzen]{Daryl Janzen}
\USask \email{daryl.janzen@usask.ca}

\author[orcid=0000-0001-8738-6011, gname=Saurabh, sname=Jha]{Saurabh W.\ Jha}
\Rutgers \email{saurabh@physics.rutgers.edu}

\author[orcid=0009-0005-1871-7856]{Ravjit Kaur} 
\UCSC \email{}

\author[orcid=0000-0001-9589-3793, gname=Michael, sname=Lundquist]{Michael ~J. Lundquist}
\Keck \email{mlundquist@keck.hawaii.edu}

\author[0000-0001-5807-7893]{Curtis McCully}
\LCO \email{}

\author[orcid=0009-0008-9693-4348, gname=Darshana, sname=Mehta]{Darshana Mehta}
\UCD \email{ddmehta@ucdavis.edu}


\author[orcid=0000-0002-7015-3446, gname=Nicol\'as, sname=Meza-Retamal]{Nicol\'as Meza-Retamal}
\UCD \email{nemezare@ucdavis.edu}

\author[orcid=0000-0003-3656-5268]{Yuan Qi Ni} \LCO \UCSB \KITP \email{}

\author[orcid=0000-0002-1092-6806]{Kishore C. Patra} 
\UCSC \email{}

\author[orcid=0000-0003-4175-4960, gname=Conor, sname=Ransome]{Conor Ransome}
\UA \email{cransome@arizona.edu}

\author[orcid=0000-0002-4022-1874, gname=Manisha, sname=Shrestha]{Manisha Shrestha}
\Monash \OzGrav \email{manisha.shrestha@monash.edu}

\author[orcid=0000-0001-5510-2424, gname=Nathan, sname=Smith]{Nathan Smith}
\UA \email{nathansmith@arizona.edu}

\author[orcid=0000-0001-8073-8731, gname=Bhagya, sname=Subrayan]{Bhagya Subrayan}
\UA \email{bsubrayan@arizona.edu}

\author[orcid=0000-0002-5748-4558]{Kirsty Taggart} 
\UCSC \email{}

\author[orcid=0000-0002-7334-2357, gname=Xiaofeng, sname=Wang]{Xiaofeng Wang} 
\Tsinghua \email{wang_xf@mail.tsinghua.edu.cn}

\author[orcid=0009-0006-7296-728X]{Kathryn Wynn}
\LCO \UCSB \email{}


\author[orcid=0009-0004-4256-1209]{Shengyu Yan}
\Tsinghua \email{yansy19@mails.tsinghua.edu.cn}

\author[0000-0002-6535-8500]{Yi Yang (杨轶)}  
\Tsinghua 
\email{yiyangtamu@gmail.com} 

\author[0000-0002-2636-6508]{Weikang Zheng} 
\Berkeley \email{}

\author[orcid=0000-0001-7410-7669]{Dan Coe} \STScI \email{}
 
\begin{abstract}
Supernova (SN)\,2025coe at a distance of $\sim$25 Mpc is the second-closest calcium-strong (CaST) transient. It was discovered at a large projected offset of $\sim$34 kpc from its potential host galaxy NGC 3277. Multiband photometry of SN\,2025coe indicates the presence of two peaks at day $\sim$2 and day $\sim$11 after explosion. Modeling the bolometric light curve, we find that the first peak can be reproduced either by shock cooling of a compact envelope ($R_\mathrm{env}$\,$\approx 6$--40 $R_{\odot}$; $M_\mathrm{env}$\,$\approx 0.1$--0.2 $M_{\odot}$) or by interaction with close-in circumstellar material (CSM; $R_{\mathrm{CSM}} \lesssim 6 \times10^{14}$ cm), or a combination of both. The second peak is dominated by radioactive decay of $^{56}$Ni ($M_{\mathrm{ej}} \approx 0.4$--0.5\,$M_{\odot}$; $M_{^{56}\mathrm{Ni}} \approx 1.4 \times 10^{-2}$\,\Mdot{}). SN\,2025coe rapidly evolves from the photospheric phase dominated by He\,I P-Cygni profiles to nebular phase spectra dominated by strong [\CaII] $\lambda \lambda$7291, 7323 and weak [\OI] $\lambda \lambda$6300, 6364 emission lines. Simultaneous line profile modeling of [\CaII] and [\OI] at nebular phases shows that an asymmetric core-collapse explosion of a low-mass ($\lesssim$3.3\,\Mdot{}) He-core progenitor can explain the observed line profiles. Alternatively, lack of local star formation at the site of the SN explosion combined with a low ejecta mass is also consistent with a thermonuclear explosion due to a low-mass hybrid He-C/O white dwarf + C/O white dwarf merger.

\end{abstract}


\section{Introduction} \label{sec:1}

Calcium-strong transients (CaSTs) are a rare category of rapidly evolving and relatively faint stellar explosions. Despite over a decade of study, their progenitor pathways remain uncertain. Proposed scenarios span both massive star core-collapse channels \citep[e.g.,][]{Kawabata10, Milisavljevic17, De21, Ertini23} and thermonuclear detonations of unusual white dwarfs \citep[WDs; e.g.,][]{Perets10, Kasliwal12, Foley15, Galbany19, Shen19, JacobsonGalan20a, Jacobson-Galan20b, Jacobson-Galan22}, with growing evidence suggesting the population may not be homogeneous.

Observationally they have been defined by  significantly stronger [\CaII] $\lambda\lambda$7291, 7324 emission compared to [\OI] $\lambda\lambda$6300, 6364 in the optically thin nebular phases \citep{Filippenko03, Kasliwal12, Valenti14b, Milisavljevic17, Gal-Yam17, Lunnan17, Jacobson-Galan22, Ertini23}. While this has led them to be often labeled ``calcium-rich,'' abundance estimates of several such supernovae (SNe) have indicated that they do not produce more Ca relative to O \citep[e.g.,][]{Milisavljevic17, Jacobson-Galan20b, Jacobson-Galan22}. Thus, we choose to adopt the ``Ca-strong" (CaST) terminology convention throughout this work \citep{Shen19}.

Typically, CaSTs are low-energy explosions \citep[$E_{k} \approx 10^{50}$ erg; peak $M_\mathrm{peak} > -16.5$ mag;][]{Taubenberger17} that produce small amounts of ejecta ($\lesssim$\,0.7\,\Mdot{}) and radioactive $^{56}$Ni ($\lesssim$\,0.1\,\Mdot{}) leading to a rapid photometric evolution. Spectroscopically, the evolution of CaSTs resembles that of stripped-envelope SNe (SESNe), but with a more rapid transition from the photospheric to the nebular phase.

Early sample studies of CaSTs have shown a strong preference for remote locations at significant offsets (as much as 150 kpc) from their host galaxies as explosion sites, suggesting that these transients arise from old stellar progenitors \citep[e.g.,][]{Perets10, Kasliwal12, Foley15, Lunnan17}. However, a growing population has confirmed heterogeneity within the class. Studies of CaSTs like iPTF15eqv \citep{Milisavljevic17}, iPTF16hgs \citep{De18}, SN\,2016hnk \citep{Galbany19, JacobsonGalan20a}, SN\,2019ehk \citep{Jacobson-Galan20b, Nakaoka21, De21}, and SN\,2021gno \citep{Jacobson-Galan22, Ertini23} suggest that a single progenitor channel cannot explain all the observed properties. Additionally, while most CaSTs share spectroscopic similarities with stripped-envelope SNe at peak luminosity (e.g., SN\,2019ehk, SN\,2021gno), a subset show peak spectra resembling those of more typical thermonuclear explosions (e.g., SN\,2016hnk). The $g-r$ color distribution of CaSTs at peak luminosity was correlated with three spectroscopic subclasses, suggesting potential differences in their progenitor systems and explosion mechanisms \citep{De20}.

High-cadence early photometry campaigns of the fast evolving CaSTs have unveiled several candidates with double-peaked optical light curves \citep[iPTF16hgs, SN\,2018lqo, SN\,2019ehk, SN\,2021gno, SN\,2021inl;][]{De18, De20, Jacobson-Galan20b, Jacobson-Galan22, Ertini23}. While the early excess suggested the presence of a compact envelope and/or circumstellar material (CSM) around the progenitor, the second peak was well explained by radioactive decay of $^{56}$Ni \citep{Jacobson-Galan22, Ertini23}. 

In this work we describe the rapid multiwavelength (ultraviolet and optical) evolution of the sixth double-peaked CaST, SN\,2025coe. At $\sim$25 Mpc, it is the second-closest CaST ever found (see Section \ref{sec:2} for details), after SN\,2019ehk at $\sim$16 Mpc. A companion paper, \cite{Kumar26} to this work will present the X-ray, near-infrared (NIR), and radio observations of SN\,2025coe. Recently, SN\,2025coe was also studied by \cite{Chen25}, where several aspects of its photometric and spectroscopic evolution were discussed. We will compare our interpretations with these results in the appropriate sections.

The discovery and observations of SN\,2025coe are presented in Sections \ref{sec:2} and \ref{sec:3}, respectively. Section \ref{sec:4} presents the extinction along the line of sight, host-galaxy properties, and local environment. The photometric and spectroscopic evolution is presented in Sections \ref{sec:5} and \ref{sec:6}, respectively. We compare the observations with physical explosion scenarios and discuss likely progenitor systems that led to SN\,2025coe in Section \ref{sec:7}. Section \ref{sec:8} summarizes our conclusions.

\section{Discovery and Classification}  \label{sec:2}
SN\,2025coe was discovered by \cite{Itagaki25} in an image taken on 2025-02-24 (yyyy-mm-dd) at 15:13:06 (UTC is used throughout this paper; MJD 60730.63) at a brightness of 17.4 mag with a clear filter. The most constraining and last-available nondetection is from the Asteroid Terrestrial-impact Last Alert System \citep[ATLAS;][]{Tonry18a, Smith20} $o$ band on 2025-02-24 00:14:24 (MJD 60730.01) down to a limit of 20.4 mag. Given a $\lesssim$1 day nondetection constraint, we approximate the explosion epoch ($t_{0}$) to be the mid-point between the first detection and last nondetection epochs at MJD = 60730.3 $\pm$ 0.3, throughout this work. The uncertainty covers the time between the first detection and the last nondetection when the explosion could have happened.

\begin{figure}
    \centering
    \includegraphics[width=0.5\textwidth]{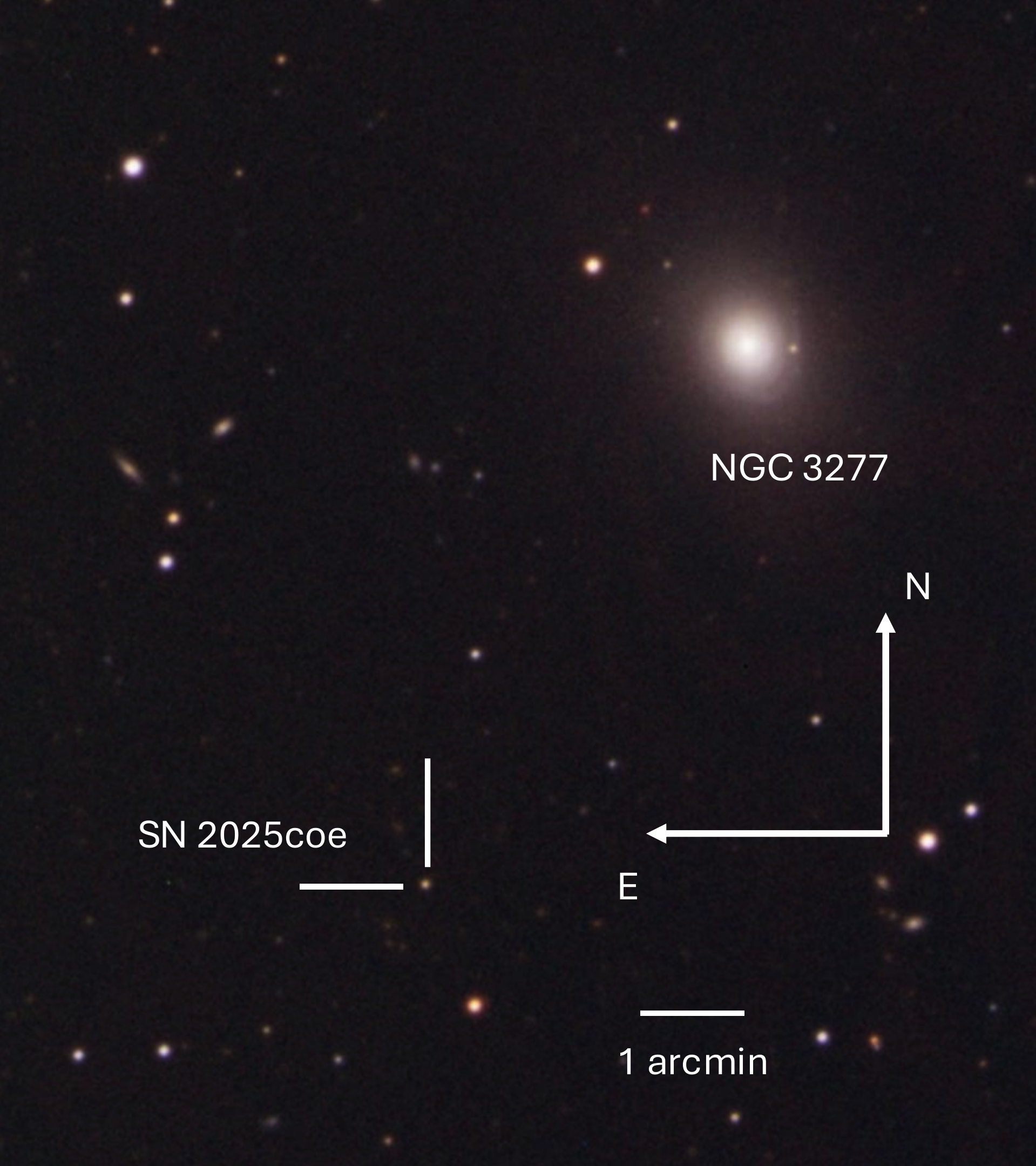}
    \caption{Three-color RGB image of the field near SN\,2025coe using Las Cumbres Observatory $g$, $r$, and $i$ filter images taken on 2025-03-21. The field of view shows the early-type spiral host galaxy NGC 3277 and SN\,2025coe at a significant projected offset of 
    $\sim$34 kpc ($\sim 5'$ from the host center). The image orientation and scale are marked. }
    \label{fig:rgb}
\end{figure}

Based on the spectroscopic redshift of $z \approx 0.0048$, SN\,2025coe is likely associated with the host galaxy NGC 3277 at a large offset. NGC 3277 is an early-type spiral galaxy with an SA(r)ab morphology \citep[][]{deVaucouleurs91}. SN\,2025coe was identified $\sim$\,5\,$\arcmin$ from the center of NGC 3277 (Figure \ref{fig:rgb}). Adopting a Tully-Fisher distance of $\mu \approx 31.99 \pm 0.80$ mag (25.1 $\pm$ 9.3 Mpc) to NGC 3277 \citep{TullyFisher88} and assuming a cosmology with H$_{0}$ = 70 km s$^{-1}$ Mpc$^{-1}$, $\Omega_{m} = 0.3$, and $\Omega_{\rm vac} = 0.7$, we estimate the projected offset of SN\,2025coe from its host to be $\sim$34 kpc. 

Based on the earliest blackbody-like spectrum, SN\,2025coe was first classified as a young SN II \citep{Andrews25Feb26}. With subsequent spectra it was reclassified as an SN Ib-peculiar \citep{Andrews25Mar07} and then finally as a Ca-strong SN \citep{Andrews25Mar18} based on the spectral similarities with Ca-strong transients SNe 2021gno and 2021inl at comparable epochs. Using SuperNova IDentification \citep[SNID;][]{Blondin_Tonry07} with the updated spectral templates of SESNe \citep{Liu_Modjaz14, Liu16, Modjaz16, Williamson23, Yesmin24}, we verified these classifications over time. This emphasizes the need for multi-epoch classifications for fast-evolving SNe.

\begin{deluxetable}{lc}
\tablecaption{SN\,2025coe: Basic Information \label{tab:sn_properties}}
\tablewidth{0pt}
\tablehead{
}
\startdata
RA (J2000) & 10\textsuperscript{h}33\textsuperscript{m}07\textsuperscript{s}.95\\
Dec. (J2000) & $+28\degr 26\arcmin 13\farcs 10$\\
Host galaxy & NGC 3277\\
Distance modulus ($\mu$)  & 31.99 $\pm$ 0.80 mag\\
Physical distance & 25.1 $\pm$ 9.3 Mpc\\
Projected offset & $\sim$34 kpc\\
Host morphology & SA(r)b\\
Redshift ($z$) & (4.8 $\pm$ 0.2) $\times$ 10$^{-3}$\\
$E(B-V)_{\rm total}$  &  0.02 mag$^\dagger$\\
Explosion epoch (MJD) & 60730.3 $\pm$ 0.3  \\
$t_{o, \mathrm{max}}$ (MJD) & 60741.2$^\star$ \\
$M_{o}^{\mathrm{peak}}$ & $-15.54 \pm 0.02$ mag$^\star$\\
\enddata
\tablenotetext{\dagger}{No host extinction assumed, and $E(B-V)_{\rm MW}$  based on \cite{SF11}.
\tablenotetext{\star}{Corresponds to peak brightness in the $o$ band due to radioactive decay.}}
\end{deluxetable}

\section{Observations and Data Reduction} \label{sec:3}

\subsection{Photometry}  \label{sec:3.1}

We obtained high-cadence $U, B, g, V, r, i$ follow-up photometry of SN\,2025coe soon after discovery until $\sim$120 days after explosion with the worldwide Las Cumbres Observatory network of 1\,m robotic telescopes \citep{Brown13}. The observations were triggered through the Global Supernova Project \citep{Howell19} and data were reduced with a PyRAF-based photometric reduction pipeline, \verb|lcogtsnpipe|\footnote{\url{https://github.com/LCOGT/lcogtsnpipe}} \citep{Valenti16}. Instrumental magnitudes are calculated using a standard point-spread-function (PSF) fitting technique in the pipeline. The apparent magnitudes of $g$, $r$, and $i$ filter images were calibrated to the APASS catalog \citep{Henden16}, while $U$, $B$, and $V$ filter images were calibrated to a Landolt catalog \citep{Landolt92} constructed using standard fields observed with the same telescope and night combinations as the SN observations. We used the PSF photometry without background subtraction as the SN is significantly offset from any other source (see Sections \ref{sec:2} and \ref{sec:5}).

We obtained ATLAS photometry in filters $c$ and $o$ with the forced photometry server \citep{Tonry18a, Smith20}. Zwicky Transient Facility (ZTF) photometry in $g$ and $r$ was obtained with the ZTF Forced Photometry Service \citep{Masci23}.

We obtained photometry of SN\,2025coe with TNOT (Tsinghua--Nanshan Optical Telescope) and TNT (Tsinghua--NAOC Telescope), 0.8\,m Ritchey-Chr\'etien telescopes located at the Nanshan Observatory in Xinjiang and the Xinglong Observatory of the National Astronomical Observatories of China (NAOC), respectively. The science frames were processed using the standard {\tt IRAF} reduction pipeline, including bias subtraction and flat-field correction. Source fluxes were measured with {\tt AutoPhot} \citep[\url{https://github.com/Astro-Sean/autophot};][]{Brennan22}, which performs automated PSF photometry.  For photometric calibration, the pipeline selected as many reference stars as possible within the field of view from the Pan-STARRS1 catalog \citep{Chambers16,Magnier20,Flewelling20}. No image subtraction was performed in the construction of the final light curves.

SN\,2025coe was followed in the ultraviolet (UV) with the {\it Neil Gehrels Swift Observatory} \citep[{\it Swift};][]{Gehrels04}. The UVOT data were reduced using the High-Energy Astrophysics Software (HEASoft\footnote{\url{https://heasarc.gsfc.nasa.gov/docs/software/heasoft/}}). We chose a source region centered at the position of the SN with an aperture radius of 3$\arcsec$ for photometry. The corresponding background was chosen from a source-free region with an aperture radius of 5$\arcsec$. We chose zeropoints for the photometry from \cite{Breeveld10} and used the latest updates to the time-dependent sensitivity corrections in 2020. SN\,2025coe was also detected in the X-rays with XRT, concurrent with the UV observations. Results from the X-ray analysis of SN\,2025coe will be presented in a companion paper \cite{Kumar26}

\subsection{Spectroscopy}  \label{sec:3.2}
We followed the optical spectral evolution of SN\,2025coe between day 1 and 116 after explosion. To minimize slit losses caused by atmospheric dispersion, the slit angle for each observation was oriented at or near the parallactic angle \citep{Filippenko82}. The complete spectral log associated with this work is presented in Appendix \ref{sec:Appendix A}.

We obtained nine optical spectra between days 1 and 38 after explosion with the FLOYDS spectrograph \citep{Brown13} mounted on the 2\,m Faulkes Telescope North (FTN) in Haleakala, Hawaii (USA). This telescope is part of the Las Cumbres Observatory network and our observations were triggered through the Global Supernova Project \citep{Howell19}.  We extracted, reduced, and calibrated the one-dimensional (1D) spectra using the standard FLOYDS reduction pipeline \citep[see][for a detailed description]{Valenti14a}.

SN\,2025coe was observed with the Binospec spectrograph \citep{Fabricant19} on the MMT Observatory at days 33, 61, and 83 after explosion. The initial data processing of flat-fielding, sky subtraction, wavelength calibration, and flux calibration was done using the Binospec IDL pipeline \citep{Kansky19}\footnote{\url{https://bitbucket.org/chil_sai/binospec/wiki/Home}}. We then used IRAF \citep{Tody86, Tody93} to extract the 1D spectrum.

We obtained two long-slit, low-resolution optical spectra of SN\,2025coe using the 2.16\,m telescope at Xinglong Observatory, Chengde, Beijing, China, equipped with the Beijing Faint Object Spectrograph and Camera (BFOSC) and a $1\farcm8$ slit with a G4 grating. All spectral data reduction was performed using the standard {\tt IRAF} pipeline, including bias subtraction and flat-field correction using halogen lamp flats, followed by 1D spectral extraction, wavelength calibration, and flux calibration.  

We obtained a spectrum of SN\,2025coe at day 41 with the Boller \& Chivens (B\&C) spectrograph at the Bok 90\,inch telescope operated by the University of Arizona and located at the Kitt Peak National Observatory. We reduced the data using a standard IRAF \citep{Tody86, Tody93} routine.

One spectrum was taken with the Goodman RED configuration on the Southern Astrophysical Research Telescope (SOAR) at day 58. The initial steps from flat-fielding, sky subtraction, and wavelength calibration were performed using the Goodman pipeline \footnote{\url{https://github.com/soar-telescope/goodman_pipeline}}. We performed the flux calibration and 1D spectral extraction using standard IRAF \citep{Tody86, Tody93} functions.

 SN\,2025coe was observed at days 2, 8, 13, 26, 44, and 57 with the Kast dual-beam spectrograph \citep{KAST} on the Lick Shane 3\,m telescope. We reduced the Kast data in a standard manner using custom data-reductions: {\sc UCSC Spectral Pipeline}\footnote{\url{https://github.com/msiebert1/UCSC\_spectral\_pipeline}} \citep{Siebert2019} and {\sc TheKastShiv}\footnote{\url{https://github.com/ishivvers/TheKastShiv}} \citep{Silverman12}.

We took spectra of SN 2025coe with the Low Resolution Imaging Spectrometer \citep[LRIS;][]{Oke95} on the 10\,m Keck-I telescope at the W. M. Keck Observatory on days 59 and 90 after explosion. These  were reduced with the \verb|LPipe| data-reduction pipeline \citep{Perley19} for steps including bias subtraction, flat-fielding, wavelength calibration, and flux calibration.

We obtained a Keck Cosmic Web Imager \citep[KCWI;][]{Morissey18} spectrum at day 116  with the 10\,m Keck-II telescope at the W. M. Keck Observatory.  This spectrum was reduced with the KCWI data-reduction pipeline \citep{Niell23} in a standard manner including bias subtraction, flat-fielding, wavelength calibration, and flux calibration, and it was extracted using QFitsViewer \citep{Ott12}.

\section{Extinction, Host, and Local Environment} \label{sec:4}

The line-of-sight extinction due to the Milky Way toward the direction of SN\,2025coe is $E(B-V)_{\rm MW}$ = 0.0229 $\pm$ 0.0005 mag \citep{SF11}. The equivalent width of Na~I~D absorption can be an empirical tracer of gas and dust \citep{Poznanski+12}. We observe no significant Na~I~D absorption features due to the Milky Way in any of the optical spectra of SN\,2025coe, consistent with low extinction.

No discernible Na~I~D absorption lines caused by the host are observed in SN\,2025coe either, as expected owing to its separation of $\sim$34 kpc from the potential host galaxy NGC 3277. Thus, throughout this work we assume the total extinction $E(B-V)_{\rm tot} = E(B-V)_{\rm MW} \approx 0.0229$\,mag, assuming the extinction law of \cite{Cardelli89} with $R_{V} = 3.1$ for multiband extinction corrections.

We identified the site of SN\,2025coe in the footprint of the DECaLS survey observations \citep{Dey19}. To estimate deep limits on any underlying host, we stack archival $g$ and $r$ DECaLS images at the SN site and perform aperture photometry with Photutils \citep{Bradley23} assuming a circular region (radius = 5 pixels). No source is detected down to 25.4\,mag in $g$ and 24.4\,mag in $r$ \citep[in AB magnitudes;][]{Oke_Gunn83}, assuming a zero-point\footnote{\url{https://www.legacysurvey.org/dr9/description/}} of 22.5\,mag \citep{Dey19}. This translates to an absolute magnitude limit of $M_{r} > -7.6$ mag and $M_{g} > -6.6$ mag, which cannot exclude the possibility of a globular cluster or ultra-faint dwarf galaxy below the detection limit \citep[e.g.,][]{Simon19} %

We identified archival {\it GALEX} near-UV (NUV) and far-UV (FUV) images of the field around NGC 3277 from the {\it GALEX} GR6 data release \citep{Bianchi14}. No significant source at the location of the SN was identified in either the NUV or FUV images. As UV brightness can be an indication of the local star-formation rate (SFR), we use \cite{Kennicutt98} resampling of the relationship from \cite{Madau98} for a \cite{Salpeter55} initial-mass function integrated from 0.1 to 100\,$M_{\odot}$,

\begin{equation}
    \frac{\mathrm{SFR}}{M_\odot\,\mathrm{yr}^{-1}} = \frac{L_\nu}{7.1 \times 10^{20}\,\mathrm{W\,Hz}^{-1}}\, ,
\label{eqn: SFR}
\end{equation}
where $L_\nu$ is the average luminosity spectral density for the FUV and NUV filters of {\it GALEX}, centered at $\lambda$1539 and $\lambda$2316, and with a bandwidth of $\lambda$616 and $\lambda$269, respectively.

We perform aperture photometry at the SN location with a circular region (radius = 5 pixels) using Photutils \citep{Bradley23}. Estimated upper limits on flux density in NUV and FUV images were extinction corrected using $R_{\mathrm{NUV}}$ = 8.20 and $R_{\mathrm{FUV}}$ = 8.24 \citep{Wyder07}. Assuming a distance of 25 Mpc, we convert these flux limits to an upper limit on average luminosity spectral density, $L_\nu$, for the {\it GALEX} NUV ($2.9 \times 10^{15}$ W Hz$^{-1}$) and FUV ($1.3 \times 10^{15}$ W Hz$^{-1}$) bands.  Using Equation \ref{eqn: SFR}, we convert $L_\nu$ to a local SFR upper limit of 4.2 $\times$ 10$^{-6}$ $M_\odot\,\mathrm{yr}^{-1}$ and 1.8 $\times$ 10$^{-6}$ $M_\odot\,\mathrm{yr}^{-1}$, associated with the nondetections in NUV and FUV, respectively. This is generally consistent with the remote location where SN\,2025coe exploded.

Low local SFRs at the site of CaSTs, significantly offset from their early-type hosts, are well known \citep[e.g.,][]{Kasliwal12, Jacobson-Galan22}. NGC 3277 is an early-type spiral host where possible active star formation has been noted \citep[e.g.,][]{Munoz-Mateos07, Edler24}. Additionally, it exhibits disturbed outskirts with shell-like structures, which could be consistent with past merger events as found in nearby galaxies with low brightness tidal features \citep{Morales18}. Such tidal debris could have low surface brightness and be significantly spread out from the main stellar body of the host \citep[e.g.,][]{Hendel15}. Post-merger star formation has also been linked with extended UV emission for early-type shell galaxies \citep[e.g.,][]{Rampazzo07}.

While no underlying host is detected in the optical and UV archival images at the site of SN\,2025coe, there are several extended faint sources around it, with the two closest being at projected offsets of $\sim$0.8 and $\sim$1.4 kpc. The closest source, WISEA J103307.52+282616.8 according to the Sloan Digital Sky Survey (SDSS) Data Release (DR) 18 catalog \footnote{\url{https://skyserver.sdss.org/dr18/}}, has an apparent brightness of $m_{g} = 23.48$ and $m_{r} = 21.95$ mag in the AB system. If this source is at a similar distance as NGC 3277, these apparent magnitudes correspond to $M_{g} = -8.51$ and $M_{g} = -10.04$ mag on the absolute-magnitude scale, making it a plausible faint dwarf galaxy candidate \citep{Simon19}. In fact, there are at least 6 such candidate dwarf galaxies around the site of SN\,2025coe within a projected offset $\lesssim$3 kpc. One caveat to note here is that most of these nearby sources (in projection) within the SDSS footprint have a large measured photometric redshift ($z \approx 0.3$--0.4), suggesting they could be background sources. However, owing to large uncertainties in the method of photometric redshift estimation, particularly for nearby faint extended objects, we cannot fully exclude the possibility that at least some of these sources are satellite galaxies of NGC 3277 and could therefore be potential birth sites for the SN progenitor.

These considerations about the ambient environment of SN\,2025coe suggest that despite a significant offset from its potential host NGC 3277, a massive-star origin cannot be entirely ruled out.

\section{Optical and UV Light Curves} \label{sec:5}
\subsection{Photometric Evolution} \label{sec:5.1}
\begin{figure*}
    \centering
    \includegraphics[width=\textwidth]{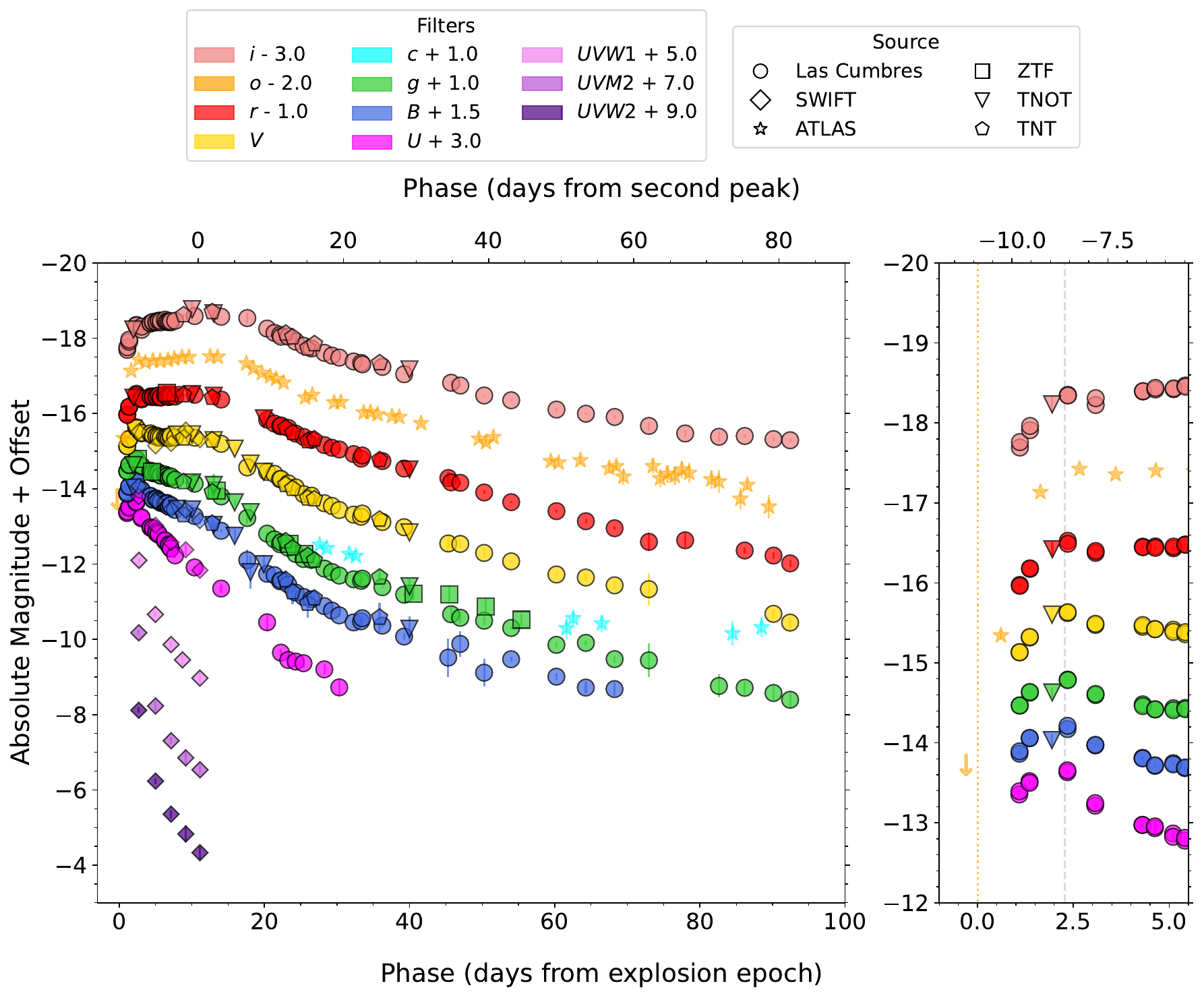}
    \caption{\textit{Left}: Multiband extinction-corrected photometry of SN\,2025coe from Las Cumbres Observatory, ZTF, ATLAS, TNOT, TNT, and {\it Swift} with respect to the epoch of explosion ($t_{0}$) and second peak ($t_{o, \mathrm{max}}$). \textit{Right}: A zoom-in view 
    of the optical light curves around the first peak (marked by a gray dashed line). The latest available nondetection from ATLAS is marked by an orange downward arrow and the estimated explosion epoch is marked by an orange dotted line.}
    \label{fig:opt_phot}
\end{figure*}

We present the optical and UV light curves of SN\,2025coe in Figure \ref{fig:opt_phot}. 
A blue excess is observed within a few days of the explosion. This is then followed by a rapid decline in brightness before rising to the second peak. The fastest decline in SN\,2025coe is observed among the bluer bands, with UV brightness decaying below {\it Swift} detection limits by day 10. To stay consistent with the convention adopted in the literature on CaSTs, we present the phase of SN\,2025coe with respect to both the explosion epoch (as discussed in Section \ref{sec:2}) and the epoch corresponding to the energy peak from radioactive decay. We estimate the second peak of the $o$-band light curve with a polynomial spline fit to be at MJD 60741.2 ($\sim$11 days after explosion) with  $M_{o}^{\mathrm{peak}} = -15.54 \pm 0.02$ mag. Basic properties of SN\,2025coe are presented in Table \ref{tab:sn_properties}.

\begin{figure*}
    \centering
    \includegraphics[width=\textwidth]{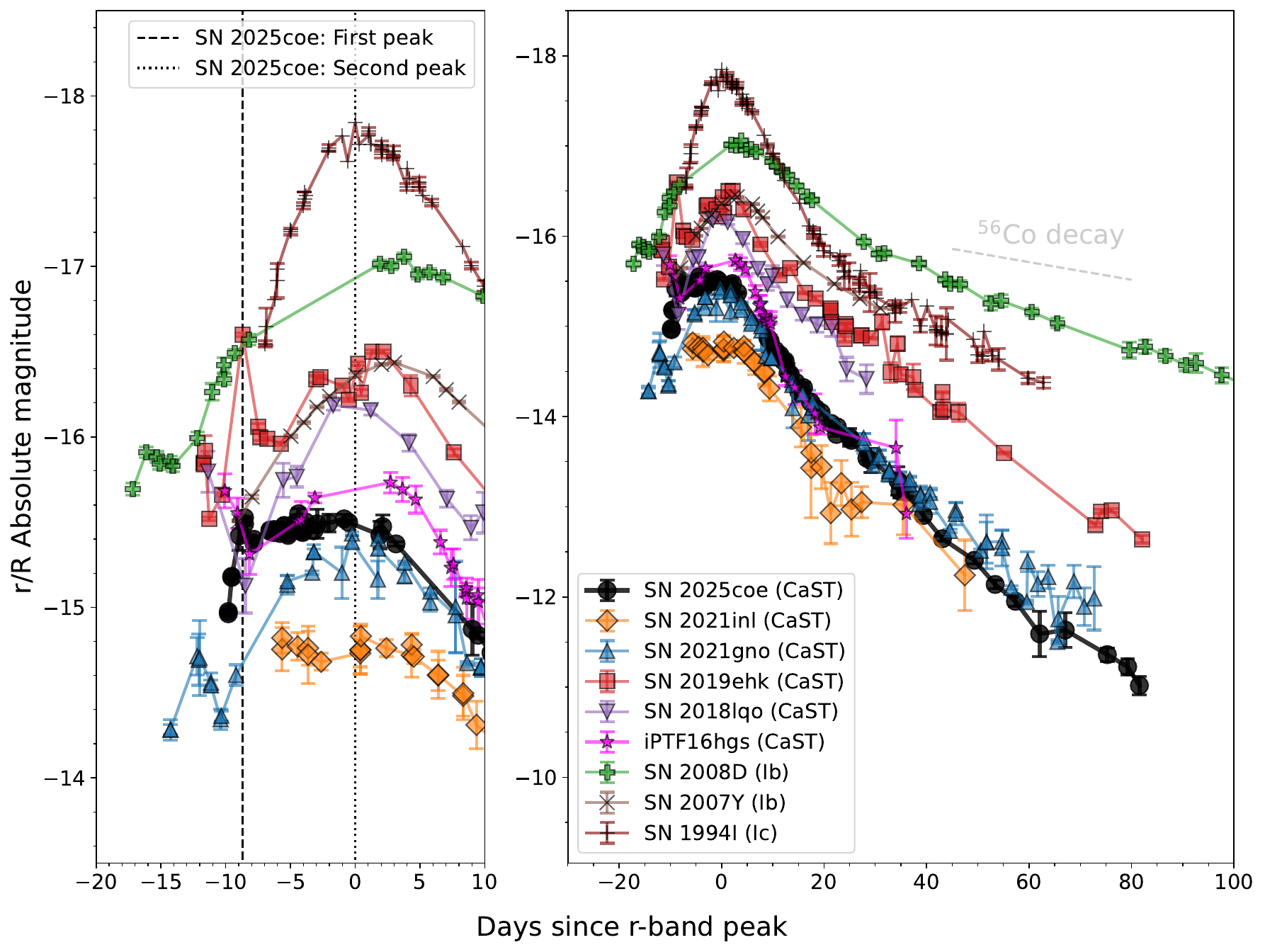}
    \caption{Extinction-corrected $r/R$ photometry comparison between SN\,2025coe and all other identified CaSTs with a double peak in their optical light curves. The first and second peaks in SN\,2025coe are marked in the zoomed-in left panel of the plot.  All double-peaked CaSTs fade faster than the expected luminosity decline from $^{56}$Co decay (dashed gray line), indicating  incomplete trapping of $\gamma$-ray photons. References for data: SN\,2021inl -- \cite{Jacobson-Galan22}; SN\,2021gno -- \cite{Jacobson-Galan22}; SN\,2019ehk -- \cite{Jacobson-Galan20b}; SN\,2018lqo -- \cite{De20}; iPTF16hgs -- \cite{De18}; SN\,2008D -- \cite{Modjaz09}; SN\,2007Y -- \cite{Stritzinger09}; SN\,1994I --\cite{Richmond96}.}
    \label{fig:phot_comp}
\end{figure*}

In Figure \ref{fig:phot_comp}, we compare the extinction-corrected $r/R$ photometry of SN\,2025coe with all other double-peaked CaSTs. Photometrically, SN\,2025coe evolves most similarly to SN\,2021gno and SN\,2021inl \citep{Jacobson-Galan22}. The peak luminosity of SN\,2025coe is consistent with other CaSTs (at $M_{r}$ $\gtrsim$ $-$16.5 mag), although it is inherently fainter than SN\,2019ehk (by $\sim$1 mag at second peak). With a quick transition to the nebular phase, all double-peaked CaSTs decline faster ($\sim$0.04--0.06 mag day$^{-1}$) than what is expected from the radioactive decay of $^{56}$Co ($\sim$0.0098 mag day$^{-1}$, for complete trapping). This suggests an incomplete trapping of $\gamma$-ray photons in the radioactive-decay process (Figure \ref{fig:phot_comp}). All double-peaked CaSTS are significantly fainter (by $\sim$2 mag) and decline faster than SN\,1994I \citep{Richmond96}, a well-studied fast-declining SN\,Ic. They also decline faster than SN\,2007Y \citep{Stritzinger09} and SN\,2008D \citep{Modjaz09}, both fast-declining SNe\,Ib. This is consistent with the general faint and fast-declining nature of all CaSTs \citep[e.g.,][]{Kasliwal12}.  

Despite some differences in brightness between SN\,2025coe, SN\,2021gno, SN\,2021inl, SN\,2018lqo, and SN\,2019ehk, they all have double peaks and reasonably similar rise times to the $^{56}$Ni-powered second peak, followed by a rapid decline during the nebular phases. \cite{Chen25} suggest the presence of a tentative third peak in SN\,2025coe at $\sim$43 days after discovery in a few photometric bands. However, we do not observe this feature from our photometric dataset, except marginally in the ZTF $g$ band, although this could be from statistical scatter as the SN is fading rapidly (Figure \ref{fig:opt_phot}). Moreover, we verified that the suggested rise in $G/Gbp/Grp$ bands from publicly available RAPAS photometry of SN\,2025coe \citep{Midavaine25} presented by \cite{Chen25} is within the uncertainty level of that dataset. Thus, we think this scatter in photometry among a few bands is statistical and not likely a true peak.

\begin{figure}
    \centering
    \includegraphics[width=0.5\textwidth]{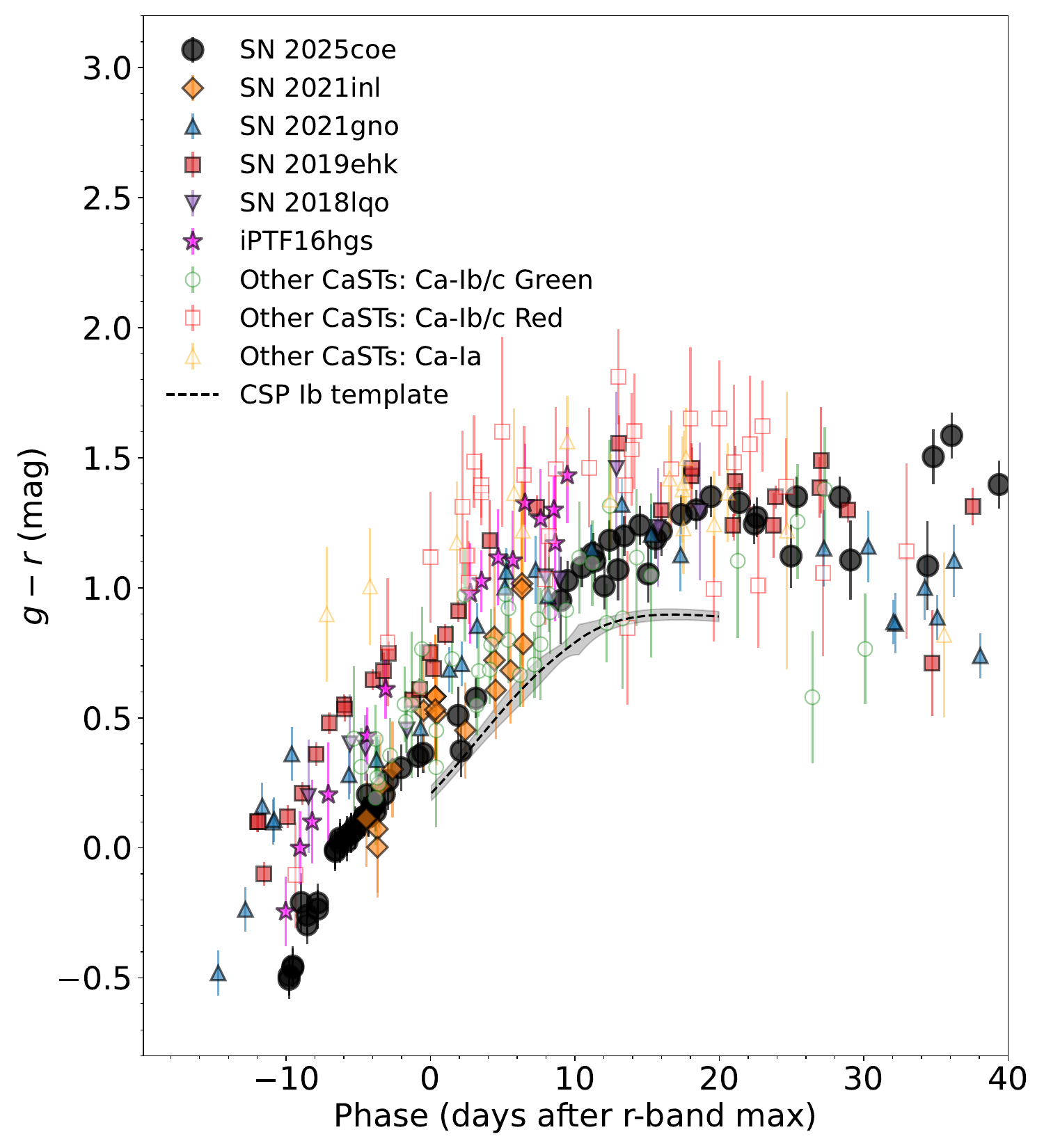}
    \caption{Extinction-corrected $g-r$ color comparison between double-peaked CaSTs: SN\,2025coe, SN\,2021inl, and SN\,2021gno \citep[][]{Jacobson-Galan22}; SN\,2019ehk  \citep[][]{Jacobson-Galan20b}, SN\,2018lqo \citep{De20}, iPTF16hgs  \citep{De18}, and other CaSTs with $gr$ photometry near peak luminosity. Data for other CaSTs are adapted from the literature \citep{Sullivan11, Kasliwal12, Valenti14, Lunnan17, De20}. The other CaSTs are marked by their membership in spectroscopic subclasses of Ca-Ib/c Green, Ca-Ib/c Red, and Ca-Ia as described by \cite{De20}.  The $g-r$ colors of SN\,2025coe and several other double-peaked CaSTs around peak match better with those of the Ca-Ib/c Green subclass. The Carnegie Supernova Project (CSP) $g-r$ color template presented by \cite{Stritzinger18} for SNe Ib is shown for comparison. Shaded region corresponds to the uncertainty in the color template values.}
    \label{fig:color}
\end{figure}

We also compare the $g-r$ colors at peak luminosity between the sample of double-peaked CaSTs with other literature-confirmed CaSTs in Figure \ref{fig:color}. The other CaSTs are color coded by their membership in the Ca-Ib/c Green, Ca-Ib/c Red, and Ca-Ia spectroscopic subclasses based on $g-r$ color at peak as discussed by \cite{De20}. Around peak brightness, while the Ca-Ib/c Red class has $g-r \approx 1.5$ mag, the Ca-Ib/c Green class has $g-r \approx 0.5$ mag \citep{De20}. There is consistency among the double-peaked sample, with the early excess being blue ($g-r$ $<$ 0 mag) followed by rapid transformation to red color ($g-r$ $>$ 1 mag) by day 20 after explosion. While the earliest color of SN\,2019ehk is apparently redder than the other double-peaked CaSTs, there is significant uncertainty in the line-of-sight extinction for its explosion site \citep{Jacobson-Galan20b, De21, Nakaoka21}. SN\,2025coe is on the blue edge of this distribution, which may indicate that SN\,2025coe has a more compact stellar envelope and/or stronger interaction with CSM than in SN\,2021gno \citep{Jacobson-Galan22, Ertini23} and SN\,2019ehk \citep{Jacobson-Galan20b, Nakaoka21}. We discuss this early blue excess in Section \ref{sec:7.1}. As the majority of CaSTs evolve similarly to SNe Ib at peak, we also show in Figure \ref{fig:color} the Carnegie Supernova Project (CSP) SNe Ib intrinsic color template of \cite{Stritzinger18}. Although the color-change behavior after peak brightness is qualitatively similar between CaSTs and SNe Ib, the colors in CaSTs are systematically shifted toward redder colors (by $\sim$0.5 mag).

\subsection{Bolometric Light-Curve Analysis} \label{sec:5.2}
\begin{figure}
    \centering
    \includegraphics[width=0.48\textwidth]{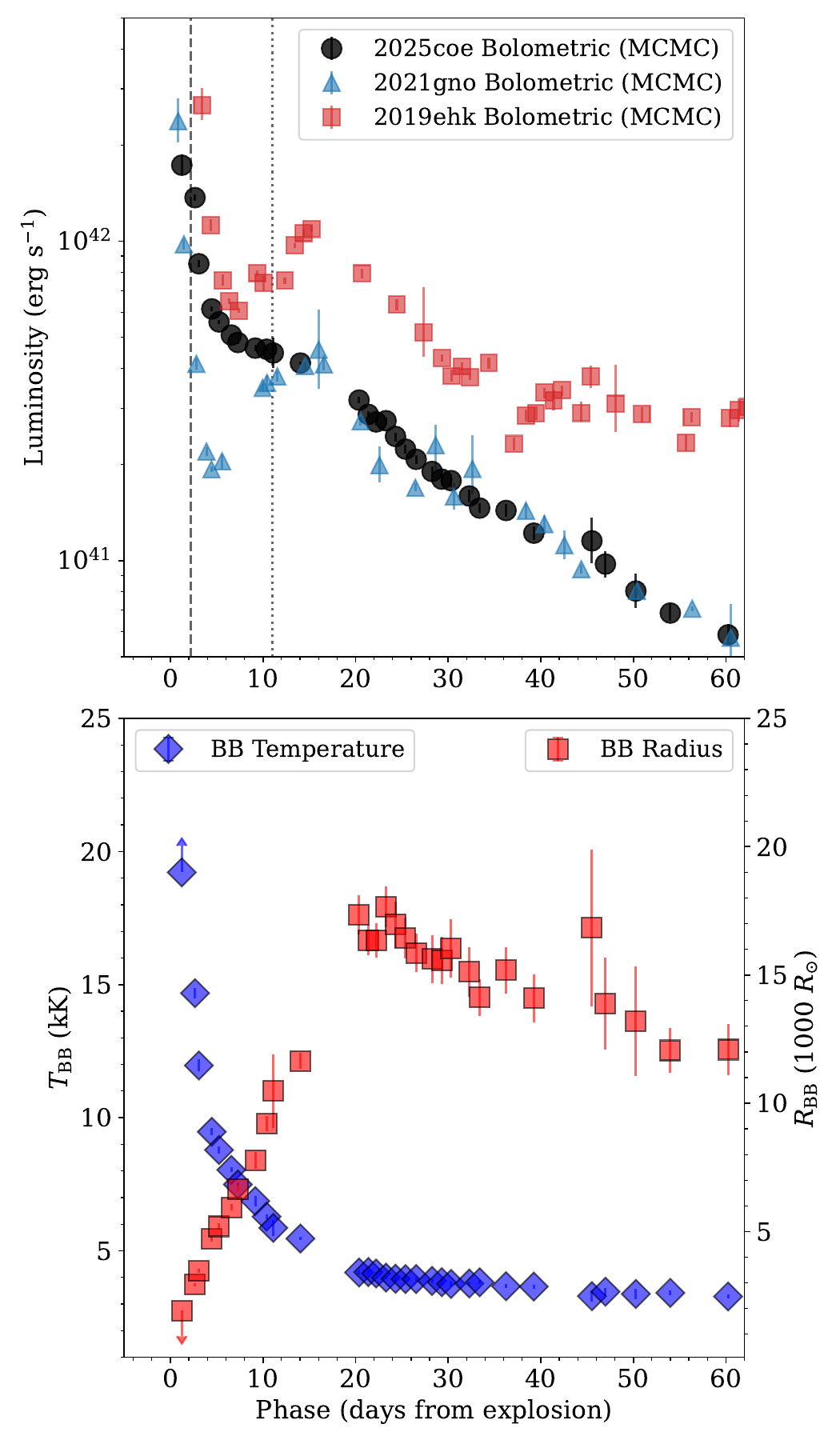}
    \caption{\textit{Top}: Bolometric luminosity from blackbody fits to the observed SED for SN\,2025coe, SN\,2021gno, and SN\,2019ehk. Optical and UV photometry for SN\,2021gno and SN\,2019ehk from \cite{Jacobson-Galan22} and \cite{Jacobson-Galan20b}, respectively. Dashed and dotted lines respectively represent the first and second peaks (estimated from optical photometry) of SN\,2025coe. \textit{Bottom}: Blackbody temperature and radius associated with the SED fit at each epoch of SN\,2025coe. As the earliest epoch lacks UV observations, the estimated radius and temperatures are upper and lower limits, respectively.}
    \label{fig:bol_temp_rad}
\end{figure}

We construct bolometric light curves of SN\,2025coe using the Light Curve Fitting package \citep{Hosseinzadeh23}. It uses a Markov Chain Monte Carlo (MCMC) routine to fit a blackbody spectrum to the observed spectral energy distribution (SED; UV through optical) at each epoch to estimate the bolometric luminosity. In Figure \ref{fig:bol_temp_rad} we plot the bolometric light curve, and the calculated photospheric temperature and blackbody radius as a function of time. We also construct bolometric light curves for SN\,2019ehk and SN\,2021gno with the same method for a consistent comparison based on optical photometric data from \cite{Jacobson-Galan20b} and \cite{Jacobson-Galan22}, and publicly available {\it Swift} UV photometry.

At the nebular phase of SN\,2025coe ($t \gtrsim 60$ days after explosion), the spectra are dominated by emission lines (e.g., [\CaII]) and deviate from a purely blackbody assumption. In addition, the continuum temperature starts to peak in the infrared and we do not have NIR photometry. Therefore, we find that computing a bolometric light curve by SED fitting becomes unreliable after day $\sim$60.
Adopting the reddening and distance from Table \ref{tab:sn_properties}, we estimate the peak bolometric luminosity to be $\sim$1.7 $\times$ 10$^{42}$ erg s$^{-1}$.

Photospheric radii and temperatures derived from blackbody fitting have significant uncertainties when fitting the SED with only optical observations at early times \citep{Arcavi22}. As the first epoch of our light curve did not have UV data and the SED likely peaks in the NUV, our estimates at this epoch are conservatively the upper radius and the lower temperature limits. The earliest inferred blackbody radius ($R_{\mathrm{BB}}$) and temperature ($T_{\mathrm{BB}}$) at $t$ = +1.2 d after explosion are $R_{\mathrm{BB}} \lesssim$\,1900\,$R_{\odot}$ and $T_{\mathrm{BB}} \gtrsim$\,19,000 K, respectively (Figure \ref{fig:bol_temp_rad}; bottom panel). At early times ($t < 20$ days), while the radius increases linearly, the temperature drops exponentially.  In the context of the third peak discussed by \cite{Chen25}, they point out a slight increase in $T_{\mathrm{BB}}$ around the third peak ($\sim$43 days after explosion). However, we argue that by this phase the ejecta are no longer optically thick (Figure \ref{fig:bol_temp_rad}, top panel; Section \ref{sec:6}) and the blackbody assumption is no longer robust. Thus, marginal changes in the inferred $T_{\mathrm{BB}}$ are less reliable. 

Under the assumption of a homologous expansion at early times ($R_{\mathrm{BB}} = R_{\star} + v_{s}t$), a large progenitor radius like that of an extended red supergiant \citep[$R_{RSG} \approx 1000$\,$R_{\odot}$;][]{Smartt15}, requires an implausibly low shock velocity ($v_{s} \approx 6000$ km s$^{-1}$) to account for a photospheric radius of $\sim 1900$\,$R_{\odot}$ at 1.2 days post-explosion. With the photospheric ejecta expansion velocity of 11,000 km s$^{-1}$ from Si~II absorption at early time (see Section \ref{sec:6.1}), we can assume a conservative lower limit on the true shock velocity to be $v_{s} \gtrsim$ \,11,000 km s$^{-1}$. This implies that SN\,2025coe had an inherently compact progenitor ($R_{\star} \lesssim$ 250\,$R_{\odot}$; likely smaller), ruling out typical RSGs that produce SNe\,II \citep{Smartt15}. Similar inferences were made for other double-peaked CaSTs, including SN\,2021gno, SN\,2021inl \citep{Jacobson-Galan22}, and SN\,2019ehk \citep{Jacobson-Galan20b}, based on the earliest blackbody radius to rule out an extended progenitor. We will discuss the modeling of the bolometric light curve in more detail in Section \ref{sec:7.1}.

\section{Optical Spectra} \label{sec:6}

\subsection{Spectral Evolution} \label{sec:6.1}

\begin{figure*}
    \centering
    \includegraphics[width=\textwidth]{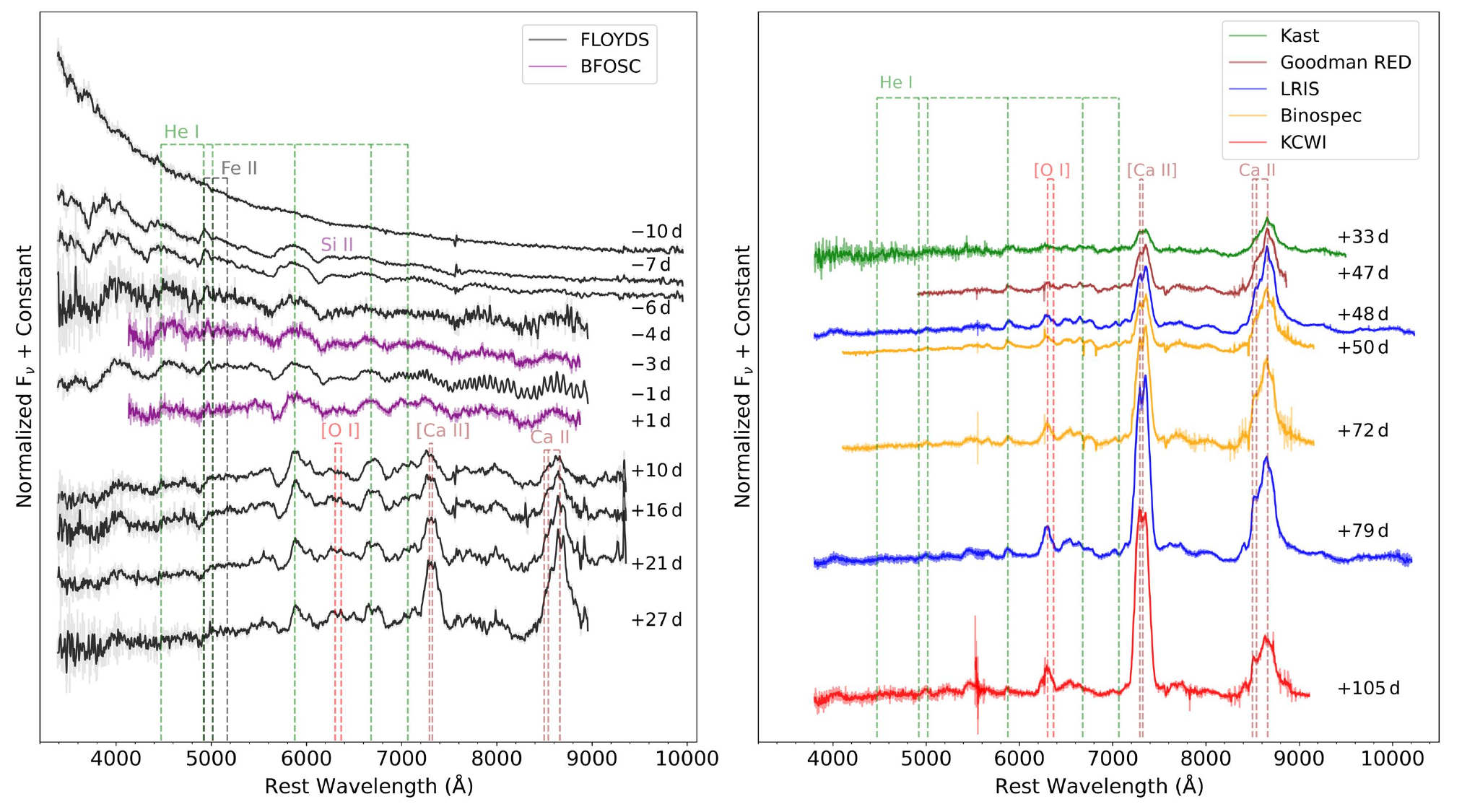}
    \caption{\textit{Left:} Optical spectral evolution of SN\,2025coe between $-10$ and 27 days from second peak. \textit{Right:} Continued optical spectral series of SN\,2025coe between 33 and 105 days. All spectra are dereddened and strong features are marked. The complete optical spectral series is described in Table \ref{tab:optical_spectralog} within Appendix \ref{sec:Appendix A}.}
    \label{fig:opt_spectra}
\end{figure*}

Our complete spectral series of SN\,2025coe is presented in Figure \ref{fig:opt_spectra}, with strong features marked. For consistency, we present phases of all spectra with respect to the second photometric peak (unless specified otherwise). At the earliest phase of $-$10\,d (at day 1 after explosion), the spectrum of SN\,2025coe looks like that of a blackbody.  By $-$7d, broad He~I features with high absorption velocities are observed. We identify broad He~I $\lambda$4471, $\lambda$5016, $\lambda$5876, $\lambda$6678, and Si~II $\lambda$6355 absorption.  The typically identified Fe~II (among other CaSTs) absorptions at $\lambda$4924, $\lambda$5018, and $\lambda$5169, are extremely weak. The early Si~II absorption feature is also typically observed in SNe\,Ib and CaSTs, but its identification can be ambiguous if there is H in the ejecta \citep[see][]{Folatelli14}. This absorption feature is no longer detected soon after +1\,d. No H features are observed at any of these early epochs, arguing against an SN\,IIb-like evolution. 

\begin{figure}
    \centering
    \includegraphics[width=0.5\textwidth]{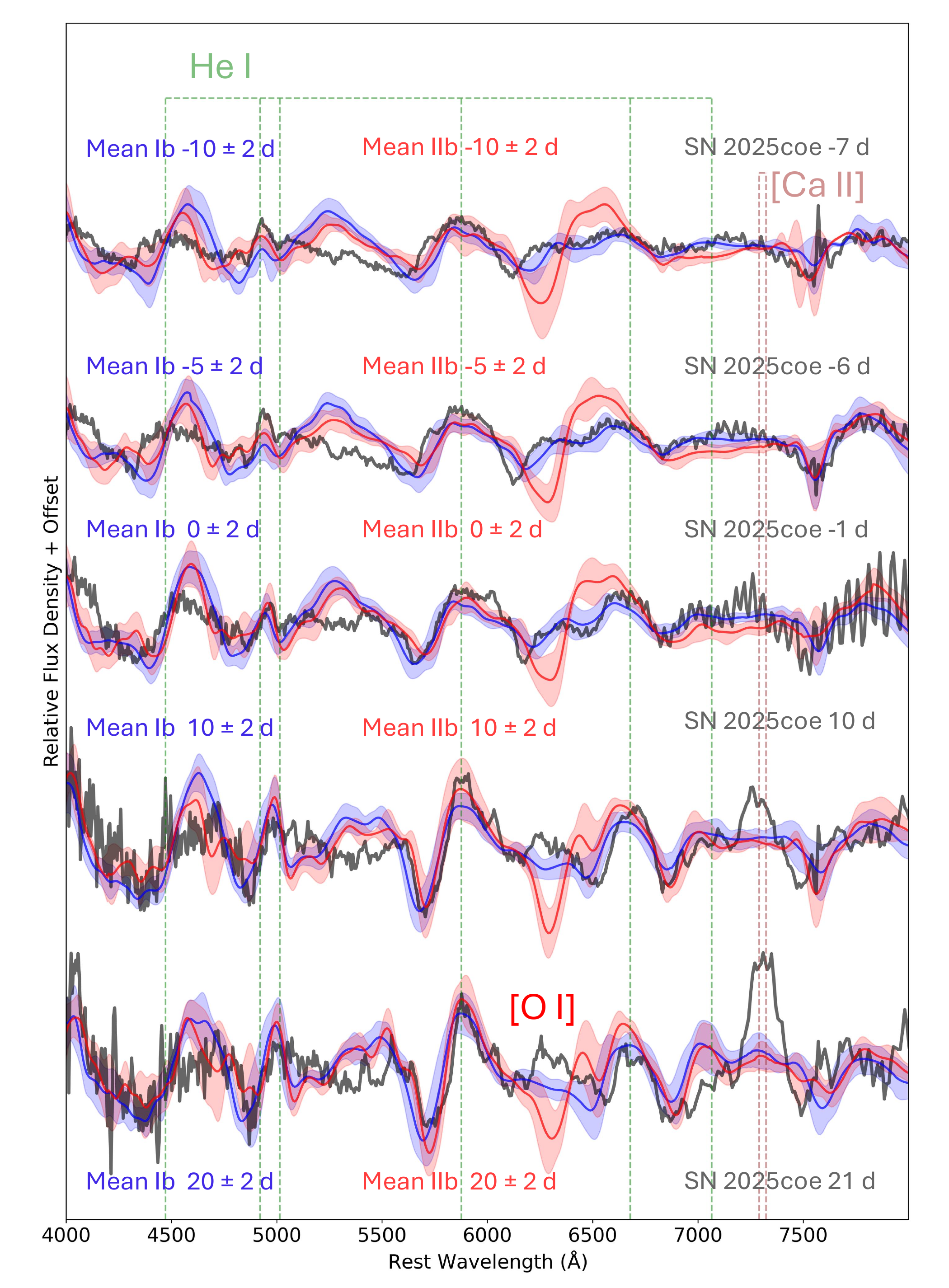}
    \caption{Comparison of flattened spectra of SN\,2025coe between $-$7 and 21 days with the mean spectra of SNe\,Ib (in blue) and IIb (in red) at similar epochs \citep{Liu16}. As early as 21 days, SN\,2025coe is more optically thin than both SNe\,Ib and IIb, showing the presence of weak [\OI] and strong [\CaII], a hallmark of the CaST population. Spectral lines of He, O, and Ca in the wavelength range marked.}
    \label{fig:template_mean_comparison}
\end{figure}

In Figure \ref{fig:template_mean_comparison} we compare the flattened spectra \citep[constructed using SNID following the procedure outlined by][]{Blondin_Tonry07} of SN\,2025coe with SN\,Ib and SN\,IIb mean spectra from the sample of \cite{Liu16}. At early times and around the second peak, the spectra of SN\,2025coe are more similar to SN\,Ib than to SN\,IIb mean spectra at comparable epochs, suggesting the absorption feature may be tentatively associated with Si II rather than H$\alpha$. However, neither template provides a perfect match to the observed spectra at these epochs and thus a robust identification of Si~II would require detailed spectral synthesis modeling, which is beyond the scope of this paper. The deviation of the [\CaII] profile in CaSTs from typical SN\,Ib evolution as early as 10 days from peak could be an observational signpost for distinguishing between these classes for future CaST classifications.

\begin{figure}
    \centering
    \includegraphics[scale=0.4]{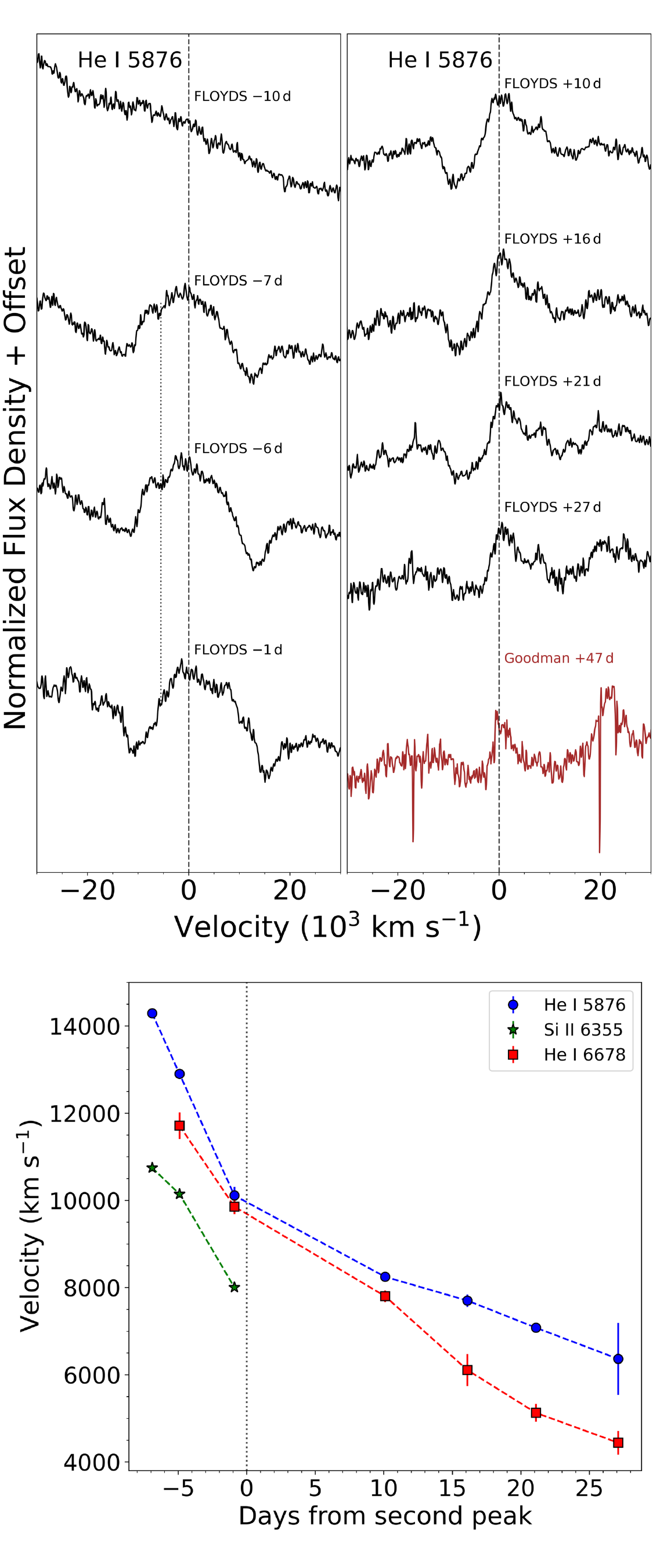}
    \caption{\textit{Top}: Evolution of the P-Cygni profile of He~I $\lambda$5876 over time plotted in velocity space between days 1 and 58. Zero velocity corresponds to the rest wavelength $\lambda$5876 (dashed gray line). The earliest spectrum is featureless and P-Cygni profiles start appearing by $-7$\,d. On days $-7$ and $-6$, a second absorption peak at a slower velocity (dotted line; $\sim$5500 km s$^{-1}$) is observed which vanishes by $-1$\,d. The overall profile changes rapidly and the absorption component vanishes by $47$ days, suggesting a quick turnaround to the optically thin phase.
    \textit{Bottom}: Estimated photospheric velocity from fitting the absorption minima of He\,I $\lambda$5876, He\,I $\lambda$6678, and Si\,II $\lambda$6355. The second-peak epoch is marked with a dashed line.}
    \label{fig:line_velocity}
\end{figure}

In the bottom panel of Figure \ref{fig:line_velocity} we present a comparison of line velocities between He~I $\lambda$5876, He~I $\lambda$6876, and Si~II $\lambda$6355. Fitting the absorption minima of He~I $\lambda$5876 gives an expansion velocity of $\sim$14,000 km s$^{-1}$ at day $-$7, which decreases to $\sim$\,6000 km s$^{-1}$ by day 27 (Figure \ref{fig:line_velocity}; bottom panel). The spectrum on $-$4\,d has a poor signal-to-noise ratio, so we do not use it to determine velocities. We estimate the photospheric velocity based on the absorption minimum of Si~II $\lambda$6355. At peak brightness, the photospheric velocity based on the Si~II absorption minimum is $\sim$8000 km s$^{-1}$. The Fe~II $\lambda$4924 absorption feature (where there could also be contributions from He~I $\lambda$5016) suggests an expanding Fe ejecta velocity of $\lesssim$4000 km s$^{-1}$, much slower than the fast-moving He. The He velocity being faster than Fe and Si suggests the presence of a fast-moving outer layer of He compared to the rest of the ejecta assuming  homologous expansion. 

On days $-$7 and $-$6, we note a secondary absorption minimum at a slower velocity ($\sim$5500 km s$^{-1}$; Figure \ref{fig:line_velocity}, top panel), which vanishes by peak (+1\,d). As the depth of the absorption is linked to the density of the foreground material, one possibility is that the transient secondary absorption is due to He ejecta clumps at a velocity of $\sim$5500 km s$^{-1}$. Clumpy distribution of He ejecta could be the consequence of an asymmetric explosion. We discuss an asymmetric explosion in the context of the progenitor of SN\,2025coe in Section \ref{sec:7.2}.
\begin{figure*}
    \centering
    \includegraphics[width=\textwidth]{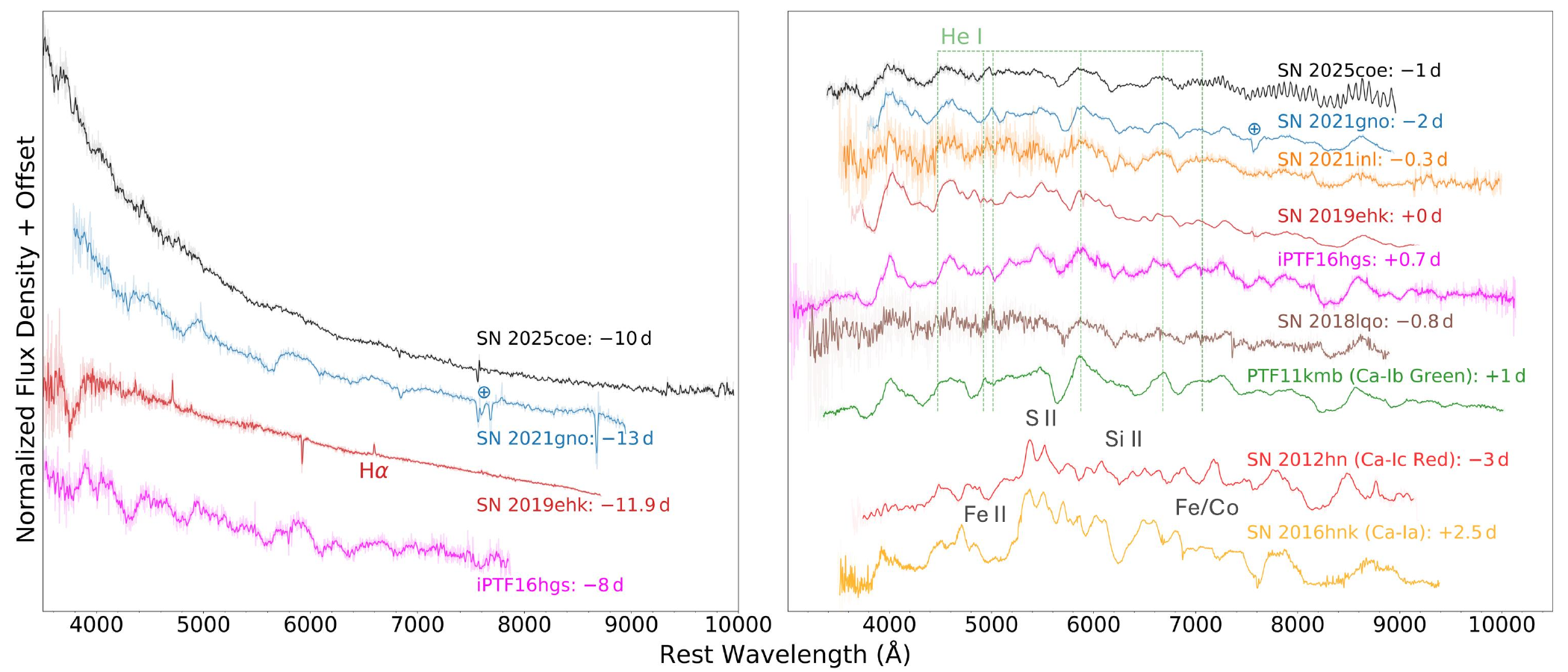}
    \caption{Comparison between extinction-corrected spectra of double-peaked CaSTs at similar epochs. \textit{Left:} CaST spectra at the earliest available epoch. The observed diversity could be due to differences in the photospheric temperature and CSM / envelope properties. \textit{Right}: Comparison of CaST spectra closest to peak brightness. The spectra between all double-peaked CaSTs look similar, with slight differences in velocity and He-line strengths. Representative CaST spectra of the Ca-Ib/c Green, Red, and Ca-Ia subclasses \citep{De20} near peak brightness are plotted for reference. SN\,2025coe and other double-peaked CaSTs most resemble the spectrum of PTF11kmb, belonging to the Ca-Ib Green subclass, and is significantly different from SN\,2012hn (Ca-Ic Red) and SN\,2016hnk (Ca-Ia). Observed He~I, S, Si~II, and Fe spectral features are marked. References for data in the plot: SN\,2021gno and SN\,2021inl -- \cite{Jacobson-Galan22}, SN\,2019ehk -- \cite{Jacobson-Galan20b}, SN\,2018lqo -- \cite{De20}, iPTF16hgs \cite{De18}, PTF11kmb -- \cite{Lunnan17}, SN\,2012hn -- \cite{Valenti14}, and SN\,2016hnk -- \cite{Galbany19}.}
    \label{fig:CaST_Ib_comparison}
\end{figure*}
We compare the spectral evolution of SN\,2025coe with other double-peaked CaSTs in Figure \ref{fig:CaST_Ib_comparison}. At the earliest epochs, SN\,2019ehk showed narrow H$\alpha$ emission suggesting the presence of H-rich CSM as opposed to broad H$\alpha$ in SNe\,IIb due to H in the expanding photosphere \citep{Jacobson-Galan20b}. No narrow lines are observed in the SN\,2025coe spectra. The diversity observed at early times between CaSTs could be due to differences in the shock-heated envelope and/or ambient CSM properties powering the luminosity at this time. 

At peak brightness, the SN\,2025coe spectrum is most similar to that of SNe\,Ib (Figure \ref{fig:template_mean_comparison}), with the lack of strong iron-group elements (IGE) typically observed in thermonuclear SNe.  Based on the spectral evolution and the presence of IGEs, Ca-strong SNe have been categorized into Ca-Ia, Ca-Ib/c Red, and Ca-Ib/c Green subclasses \citep{De20}. SN\,2025coe and all other double-peaked CaSTs show significant spectral similarities at peak light with the peak spectrum of Ca-Ib Green PTF11kmb (Figure \ref{fig:CaST_Ib_comparison}). This is consistent with peak $g-r$ colors of SN\,2025coe and most other double-peaked CaSTs (Figure \ref{fig:color}). \cite{De20} suggested low-efficiency burning scenarios like shell-only detonations or deflagrations of low-mass WDs to explain some of the Ca-Ib/c Green subclass observations. We discuss potential progenitor scenarios for SN\,2025coe in Section \ref{sec:7.2}.
 
\begin{figure}
    \centering
    \includegraphics[width=0.5\textwidth]{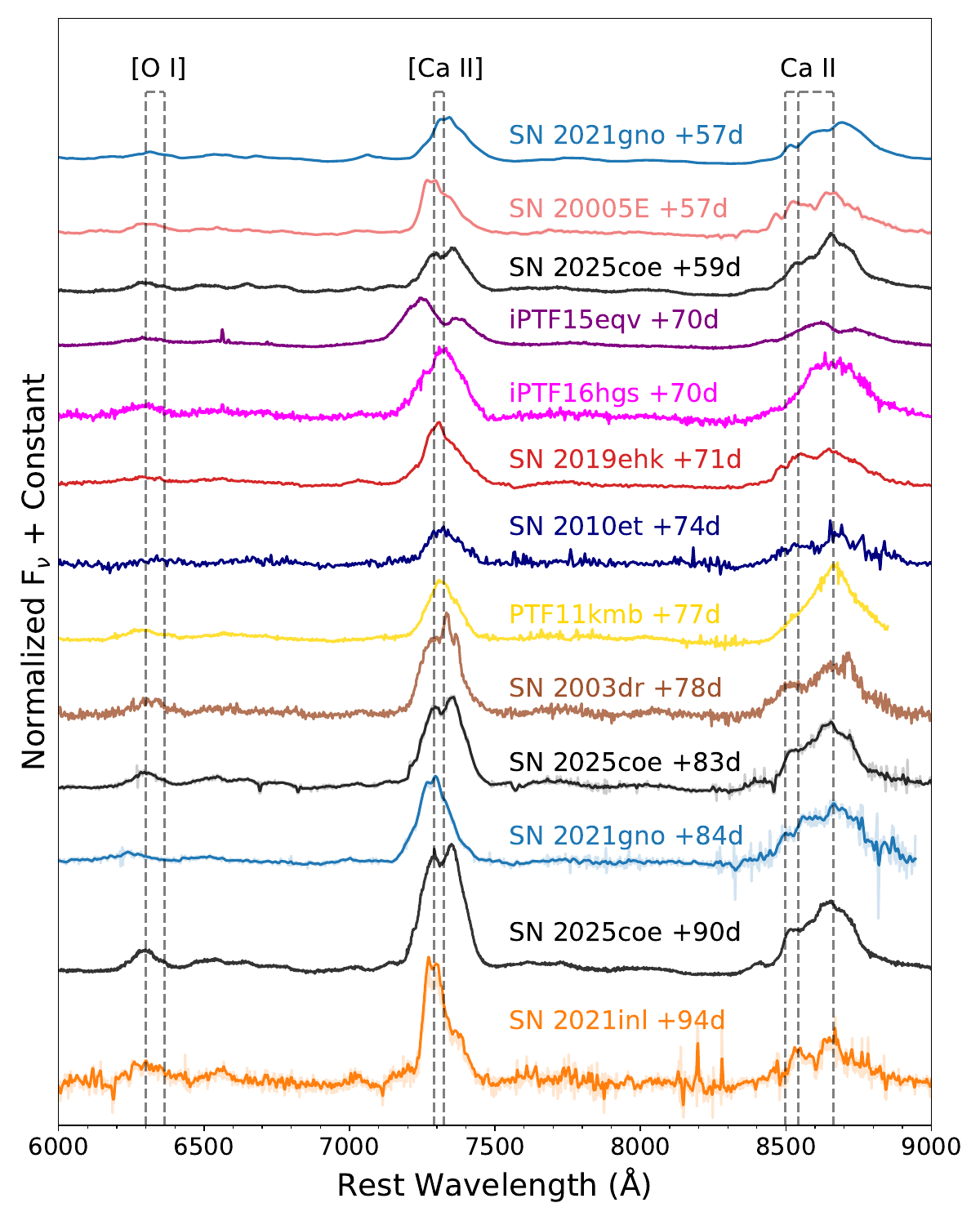}
    \caption{\textit{Top:} Comparison of the nebular spectra of SN\,2025coe with several confirmed CaSTs at comparable phases with respect to their explosion / discovery epochs. The spectra are marked by the presence of strong [\CaII] and weak [\OI]. SN\,2025coe and iPTF15eqv show the most prominent distinct double-component [\CaII]. References for data in the plot: SN\,2003dr -- \cite{Shivvers17}; SN\,2010et -- \cite{Kasliwal12}; PTF11kmb -- \cite{Lunnan17}; iPTF15eqv -- \cite{Milisavljevic17}; iPTF16hgs -- \citep{De18}; SN\,2019ehk -- \cite{Jacobson-Galan20b}; SN\,2021gno, SN\,2021inl -- \cite{Jacobson-Galan22}.}
    \label{fig:nebular}
\end{figure}


\subsection{Nebular Spectra: [\CaII]/[\OI] Ratios} \label{sec:6.2}

SN\,2025coe rapidly transitions to the nebular phase and the spectra are characterized by the  presence of strong [\CaII], which starts appearing as early as 21 days after explosion. Like several other CaSTs, SN\,2025coe shows a clear detection of [\OI] lines in the nebular spectra. A comparison of nebular spectra of double-peaked CaSTs and SN\,2025coe is shown in the top panel of Figure \ref{fig:nebular}. Unlike most CaSTs in the literature and members of the double-peaked (in light curve) subcategory of CaSTs, the nebular spectra of SN\,2025coe show two clearly distinguished components in the [\CaII] profile. iPTF15eqv is another known CaST where a similar two-component shape in [\CaII] was noted \citep{Milisavljevic17}. On the other hand, the [\OI] profile at these epochs only exhibits a single-component doublet (Figure \ref{fig:nebular}; bottom panel). We discuss a physical model to simultaneously fit the two-component [\CaII] doublet and single-component [\OI] doublet in Section \ref{sec:7.2}.

In the nebular phase between days 48 and 105, we measure the integrated and continuum-subtracted [\CaII] and [\OI] line fluxes. The estimated [\CaII]/[\OI] flux ratio for SN\,2025coe stays $\gtrsim 10$ across nebular epochs, assuming both line profiles are contributed entirely by Ca and O species, respectively (we discuss this assumption further in Section \ref{sec:7.2}). These ratios are also generally consistent with those of other double-peaked CaSTs \citep{Jacobson-Galan22}.


Following the nebular analysis outlined by \cite{Jacobson-Galan20b} and \cite{Jacobson-Galan22}, we estimate the abundance of Ca and O in SN\,2025coe by relating the observed luminosities of [\CaII] and [\OI] to the populations of the excited states, ion number densities, and the Einstein A coefficients of each ion. At densities higher than 10$^{7}$ cm$^{-3}$, this can be expressed as 
\begin{equation}
    L_{\mathrm{[O~I]}} = n_{\mathrm{O~I}} A_{\mathrm{O~I}} h\nu_{\mathrm{O~I}} (5/14) e^{-22000/T}\, 
    \label{eqn:OI}
\end{equation}
\begin{equation}
        L_{\mathrm{[Ca~II]}} = n_{\mathrm{Ca~II}} A_{\mathrm{Ca~II}} h\nu_{\mathrm{Ca~II}} (5/14) e^{-19700/T}\, 
        \label{eqn:CaII}
\end{equation}
where $h\nu$ corresponds to the photon energy ($h\nu_{\mathrm{O~I}}$ = 3.16 $\times$ 10$^{-12}$ erg; $h\nu_{\mathrm{Ca~II}}$ = 2.72 $\times$ 10$^{-12}$ erg), $n$ is the ion number density, and the Einstein A coefficients  $A_{\mathrm{Ca~II}}$ = 2.6 s$^{-1}$ and $A_{\mathrm{O~I}}$ = 340 $A_{\mathrm{Ca~II}}$. The exponentials are Boltzmann factors ($T$ in K units), and the numerical factors are statistical weights.

Converting the observed [\CaII] and [\OI] nebular line fluxes of SN\,2025coe at day 105 (the latest available spectrum) into luminosities assuming the distance of $\sim$\,25.1 Mpc, we get $L_{\mathrm{[Ca~II]}}$ = 2.3 $\times$ 10$^{39}$ erg~s$^{-1}$ and $L_{\mathrm{[O~I]}}$ = 1.9 $\times$ 10$^{38}$~erg~s$^{-1}$. Considering a typical excitation temperature range of 5000--10,000 K in Equations \ref{eqn:OI} and \ref{eqn:CaII}, we can estimate the following masses for O and Ca, respectively: $M(O) \approx  0.07$--0.6\,\Mdot{} and $M(Ca) \approx (2$--9) $\times 10^{-3}$ \,\Mdot{}. The lower mass limit corresponds to higher temperatures and vice versa. The ion number densities of [\OI] and [\CaII] are converted into mass through multiplication by the atomic masses of O and Ca, respectively. 

Like previous results from iPTF15eqv \citep{Milisavljevic17}, SN\,2021gno and SN\,2021inl \citep{Jacobson-Galan22}, these mass estimates further confirm that the strength of Ca in SN\,2025coe does not indicate that these explosions produce more calcium relative to oxygen in an absolute sense. Instead, the strength of calcium emission likely arises from the degree of mixing, ionization, and excitation conditions in the ejecta. One important caveat to note in these calculations is that they are based on a spectrum of SN\,2025coe at day 105 from peak. At later times SN\,2025coe may become further optically thin, revealing more of its ejecta. Thus, the estimated masses here are lower limits, although additional errors are also propagated through the uncertainties on the distance to SN\,2025coe.

\section{Discussion} \label{sec:7}
\subsection{Explosion Properties}\label{sec:7.1}
\subsubsection{Bolometric Light-Curve Modeling}\label{sec:7.1.1}
\begin{figure*}
    \centering
    \includegraphics[width=\textwidth]{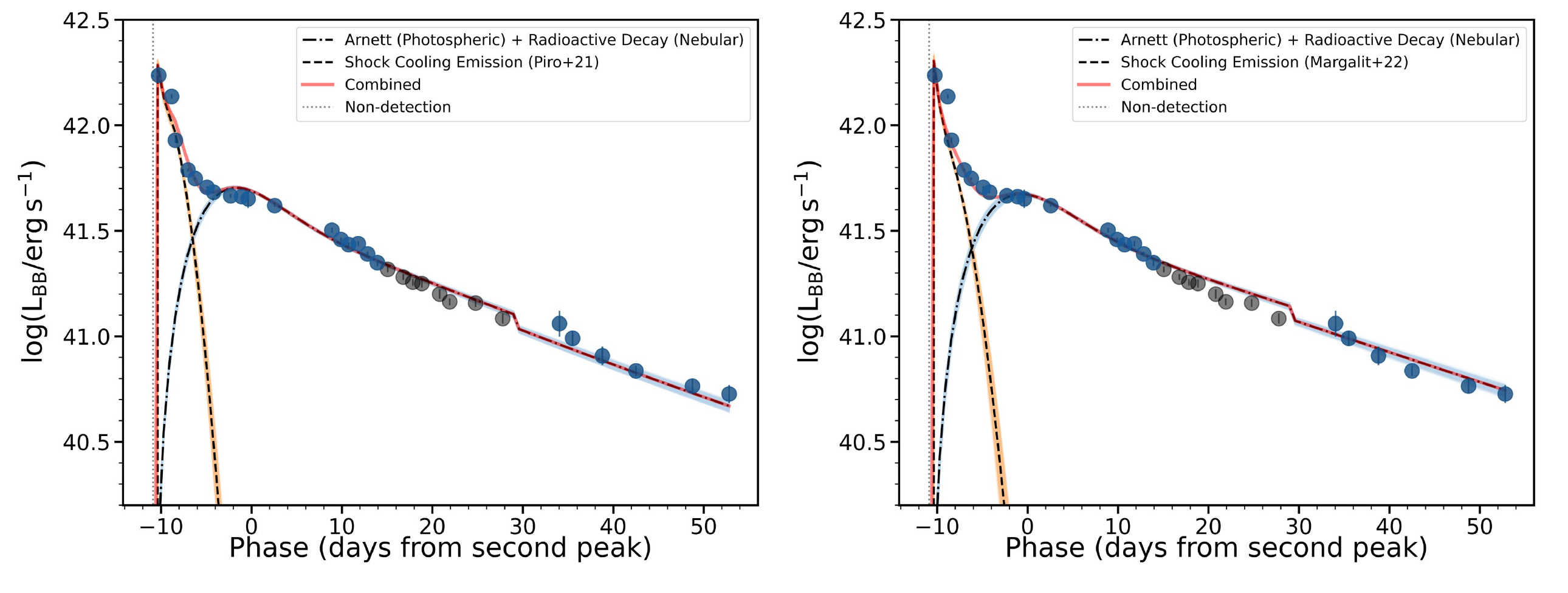}
    \caption{The bolometric light curve of SN\,2025coe is modeled with a combination of power from shock-cooling emission and radioactivity using an MCMC routine. For the shock-cooling emission, we consider the two-zone envelope model of \cite{Piro21} (left) and an analytic solution described by \cite{Margalit22} (right). In both panels, the orange and blue light curves are shock cooling and radioactivity models (500 random posterior draws), and the dotted and dash-dotted lines represent their medians, respectively. The combined median posterior light curves are shown in solid red. The strong nondetection limit for SN\,2025coe is marked by dotted lines. Across the two independent models, we find that the early excess in SN\,2025coe can be modeled with a compact envelope having a radius of $\sim$6--40\, $R_{\odot}$ and a mass of $\sim$0.1--0.2\, $M_{\odot}$. Best-fit parameters and their covariances are shown in Appendix \ref{sec:Appendix B} (Table \ref{tab:bolometric_fitting}; Figure \ref{fig:corner_plots})}
    \label{fig:bol_fitting}
\end{figure*}

SN\,2025coe stays blue for the first week after explosion (Figure \ref{fig:color}), showing a rapid decline in the UV and blue bands (Figure \ref{fig:opt_phot}). The double-peaked bolometric light curve of SN\,2025coe suggests the presence of more than one power source. The first peak and its duration are similar to the envelope-cooling emission typically observed in other double-peaked CaSTs \citep{Jacobson-Galan20b, Jacobson-Galan22, Ertini23} and CCSNe with an extended envelope \citep[e.g., IIb;][]{Targtalia+17, Crawford25, Subrayan+25}. However, in the case of SNe\,IIb, the early envelope is primarily H, left over from binary companion stripping. No signs of early H are observed in the spectra of SN\,2025coe (Figure \ref{fig:template_mean_comparison}), suggesting the envelope could instead be He-rich. This is consistent with fast-moving He absorbing optical radiation at early epochs (Figure \ref{fig:line_velocity}). As we saw in Section \ref{sec:5.2}, SN\,2025coe likely originated from a compact progenitor, making the scenario of a compact envelope plausible.

To constrain the properties of the envelope that can be inferred from the bolometric light curve, we adopt the formalism from \cite{Piro21}. In this model, extended material with mass $M_\mathrm{env}$ at radius $R_\mathrm{env}$ is imparted an energy $E_\mathrm{env}$ as the shock propagates through it. Building on the work of \cite{Chevalier_Soker89}, the extended material is divided into two zones: an outer lower-density region with a steep radial dependence ($\rho_{\mathrm{out}} \propto r^{-10}$) and a higher density inner region with shallower radial dependence ($\rho_{\mathrm{in}} \propto r^{-1}$). Homologous expansion is assumed and the luminosity due to a cooling envelope is found to be proportional to the initial envelope radius.  

As the envelope properties (mass, radius) are generally strongly dependent on model assumptions about the density structures, we also used the independent analytical prescription of shock-cooling envelope emission models described by \cite{Margalit22} for our fits. These models differ from those of \cite{Piro21} in the assumed shocked-CSM density distribution and the treatment of radiative diffusion. While \cite{Piro21} uses a two-zone broken power-law density structure, \cite{Margalit22} assume a sharp truncation of the density profile at $r = R_{0}$, where $R_{0}$ is the outer extent of material that can interact with the ejecta. 

The second peak in the double-peaked light curves of CaSTs has been suggested to be powered by the radioactive decay of $^{56}$Ni. Thus, to determine the physical parameters such as ejecta mass ($M_{\mathrm{ej}}$) and radioactive nickel mass ($M_{^{56}\mathrm{Ni}}$), we model the bolometric light curve of SN\,2025coe with a combination of photospheric and nebular models. In the photospheric phase (phase $< 30$~days after explosion), the light curve is controlled by the photon diffusion time which is a function of $M_{\mathrm{ej}}$, the ejecta velocity, and the opacity \citep{Arnett82}. Assuming that around the radioactive-decay-powered peak of SN\,2025coe the rise time is equal to the photon diffusion time, we can estimate $M_{\mathrm{ej}}$ and $M_{^{56}\mathrm{Ni}}$. We fix the optical opacity ($\kappa_{\mathrm{opt}}$) to be 0.1 cm$^{2}$ g$^{-1}$. For the nebular phase, the decay rate in the bolometric light curve is consistent with other double-peaked CaSTs. We adopt the analytical formalism described by \cite{Valenti08} where the modeling self-consistently implements the possibility of incomplete $\gamma$-ray trapping.

In Figure \ref{fig:bol_fitting}, we present the combined (shock-cooling emission + radioactivity) best-fit models and sampled posterior light curves with both these formalisms of shock-cooling emission. For our model fitting, we implement an ensemble sampler with \verb|emcee|, a Python-based affine invariant MCMC application \citep{Foreman-Mackey13}. Across the two independent models, we find considerable agreement in the inferred envelope properties. We find that a compact envelope of radius $R_{e} \approx 6$--40\, $R_{\odot}$ and mass $M_\mathrm{env} \approx 0.1$--0.2\,$M_{\odot}$ can explain the early bolometric excess. From the second peak and subsequent decline, we infer $M_{\mathrm{ej}} \approx 0.4$--0.5\,$M_{\odot}$ and  $M_{^{56}\mathrm{Ni}} \approx 1.4 \times 10^{-2}$\, \Mdot{}. Assuming a homogeneous density of the ejecta, the degeneracy between ejecta mass $M_{\mathrm{ej}}$ and kinetic energy $E_{\mathrm{k}}$ can be broken with information on the photospheric velocity ($v_{\mathrm{ph}}$) from spectroscopy \citep{Arnett82}. With $v_{\mathrm{ph}} \approx 8000$ km s$^{-1}$ based on the absorption minimum of Si~II\,$\lambda$6355 at peak brightness (see Section \ref{sec:6.1}), we estimate $E_{\mathrm{k}} \approx (0.4$--0.5) $\times 10^{51}$ erg. The low ejecta and nickel masses of SN\,2025coe with a low explosion energy are consistent with other CaSTs. Within uncertainties, the estimated explosion epochs in our combined fits are consistent with the strong nondetection limit of SN\,2025coe (see Section~\ref{sec:2}).

In both independent two-component fits, validity of the shock-cooling phase is ensured as the early component fits the data at $t \lesssim 4$ days (Figure \ref{fig:bol_fitting}). During this phase, the photospheric blackbody temperature is $\gtrsim 9000$\,K ($\sim$0.8\,eV; Figure \ref{fig:bol_temp_rad}), which is consistent with the general regime of the temperatures described by the shock-cooling envelope models. All posterior distributions for the fitted parameters are unimodal, although we observe a degeneracy between $R_\mathrm{env}$ and $M_\mathrm{env}$ as expected \citep{Piro21, Margalit22}. The best estimates of the fitted parameters and their covariances are presented in Appendix \ref{sec:Appendix B} (Table \ref{tab:bolometric_fitting}; Figure \ref{fig:corner_plots}). As these model fits involve significant simplifications and assumptions on the envelope density profiles, the fitted parameters should be considered as only order-of-magnitude estimates.

The shock-cooling parameter space of the five double-peaked CaSTs  \citep{De18, Jacobson-Galan20b, Jacobson-Galan22}, with SN\,2025coe is generally consistent. We confirm that on average, the early blue excess in these objects can be modeled with shock cooling from extended material within a radius of $\sim$5--120\,\Rdot{} and an envelope mass of $\sim$0.05--0.2\,\Mdot{}. Compared to shock-cooling model parameters presented in the literature, CaSTs show a similar extended mass to fast-rising events such as the ultrastripped SN\,2019dge \citep{Yao20} and SNe IIb \citep[e.g., IIb SN\,2016gkg, SN\,2024uwq][]{Targtalia+17, Subrayan+25}, though the latter typically exhibit a larger envelope radius consisting of H. 

The ``best-fit" shock-cooling emission parameters across different works are dependent on the model assumptions and thus need caution during a direct comparison.  \citet{Chen25}  fit the bolometric light curve of SN\,2025coe with the older models of \cite{Piro15}. While \cite{Piro15} assumes a single zone of uniform density profile for the extended envelope, \cite{Piro21} updates this to a two-zone model that subdivides the extended envelope into a compact, dense core and a more diffuse outer region. A significantly smaller envelope mass $M_\mathrm{env} \approx 1.4  \times 10^{-3}$\, $M_{\odot}$ was inferred in their work with this model. This difference in envelope mass is primarily due to the degeneracy between envelope radius and mass: for a given light-curve shape, larger envelope radii with larger masses can mimic smaller radius and lower mass scenarios. With fewer assumptions in \cite{Piro21} that constrain these degeneracies, a wider range of envelope parameter space is explored. 
 
The early blue emission could also be due to interaction with a close-in CSM distribution. X-ray and radio observations of SN\,2025coe  will be presented in a companion paper to our work \citep{Kumar26}. They find that SN\,2025coe shows {\it Swift} X-ray detections  2--8 days after explosion before decaying below detection thresholds. By estimating a spherical volume of the interacting CSM responsible for producing X-rays, they find the radial extent of the CSM to be at least $2.1 \times 10^{15}$\,cm based on the last X-ray detection (assuming a shock velocity of $3 \times 10^{4}$ km s$^{-1}$) with a total mass of $\sim$0.1\,\Mdot{}. To create this much CSM that only extends to $\sim 10^{15}$ cm, the progenitor of SN\,2025coe must have lost mass in the last months to years prior to the explosion or have been surrounded by a dense medium created by some other process. At $\sim$8 days, the blackbody photospheric radius from our SED fitting is around 8000\,$R_{\odot}$ ($5.6 \times 10^{14}$\,cm), and thus roughly consistent with where the X-rays are estimated to be coming from at this phase. Radio nondetections after 10 days as discussed by \cite{Kumar26} also suggest lack of an extended CSM distribution. 

The envelope mass ($\sim$0.1--0.2\,\Mdot{}) inferred assuming shock-envelope cooling and the independently inferred CSM mass ($\sim$0.1\,\Mdot{}) from the early X-ray observations \citep{Kumar26} are consistent with each other. Our inferred envelope mass is closer to the X-ray-based CSM estimate compared to the values presented by \cite{Chen25}. Whether it is actually a gravitationally bound compact envelope or close-in but unbound CSM (or a combination of both), are degenerate scenarios which cannot be distinguished with our data. To the nearest order of magnitude, the estimated envelope radius is also consistent with a compact progenitor as we interpreted from blackbody fits at early times (see Section \ref{sec:5.2}). Thus, the source of the observed early blue excess is likely a combination of a compact envelope associated with the progenitor and an ambient compact CSM around the progenitor due to violent mass loss before explosion. The lack of prolonged emission from interaction in both cases points to a low-mass compact progenitor.

\subsubsection{Probing Core-collapse Explosion Asymmetries with Nebular Spectra} \label{sec:7.1.2}
\begin{figure}
    \centering
    \includegraphics[width=0.45\textwidth]{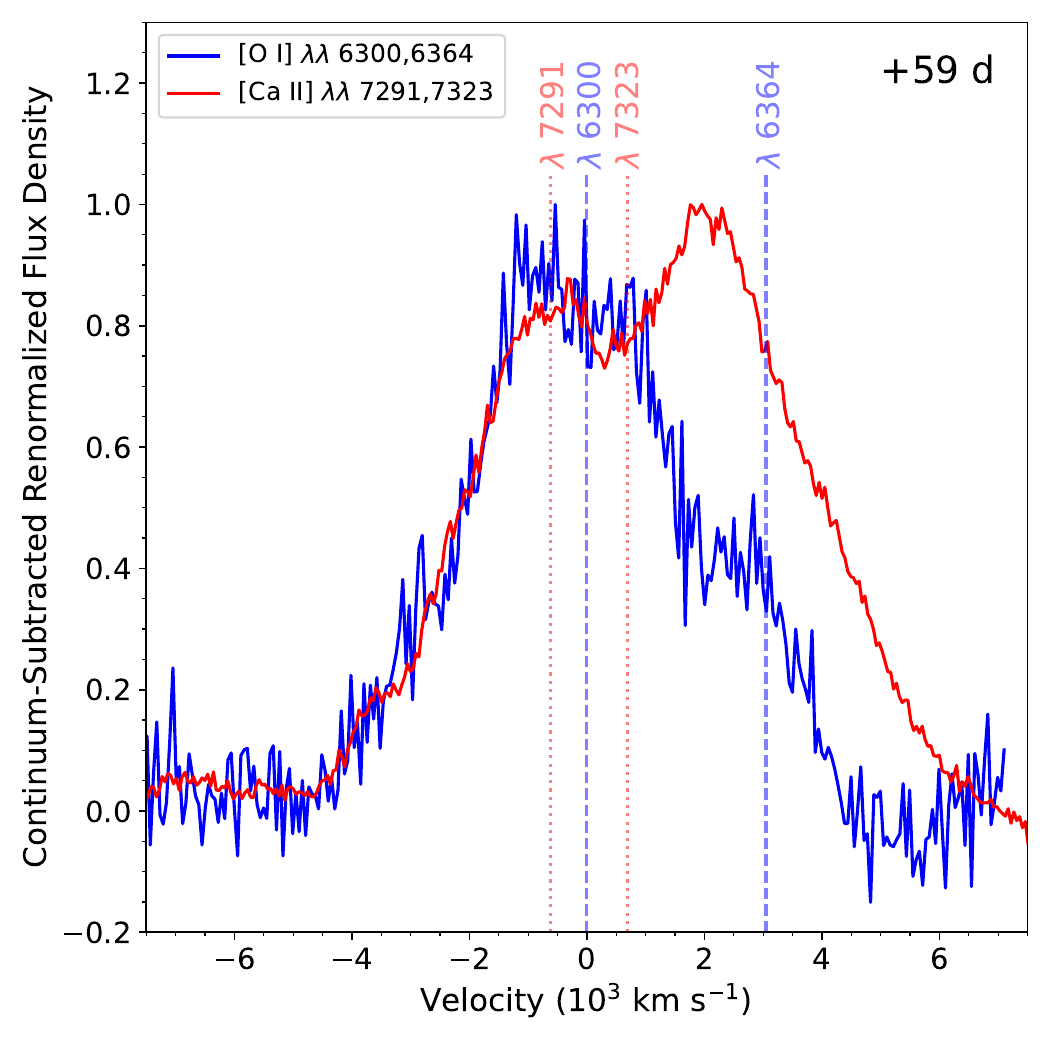}
    \caption{Comparison between continuum-subtracted and renormalized [\OI] and [\CaII] line profiles  in the nebular spectrum at 59 days after explosion. Zero velocities are with respect to $\lambda$6300 and $\lambda$7306 for [\OI] and [\CaII], respectively.}
    \label{fig:OI_CaII_profile}
\end{figure}
\begin{figure*}
    \centering
    \includegraphics[scale=0.42]{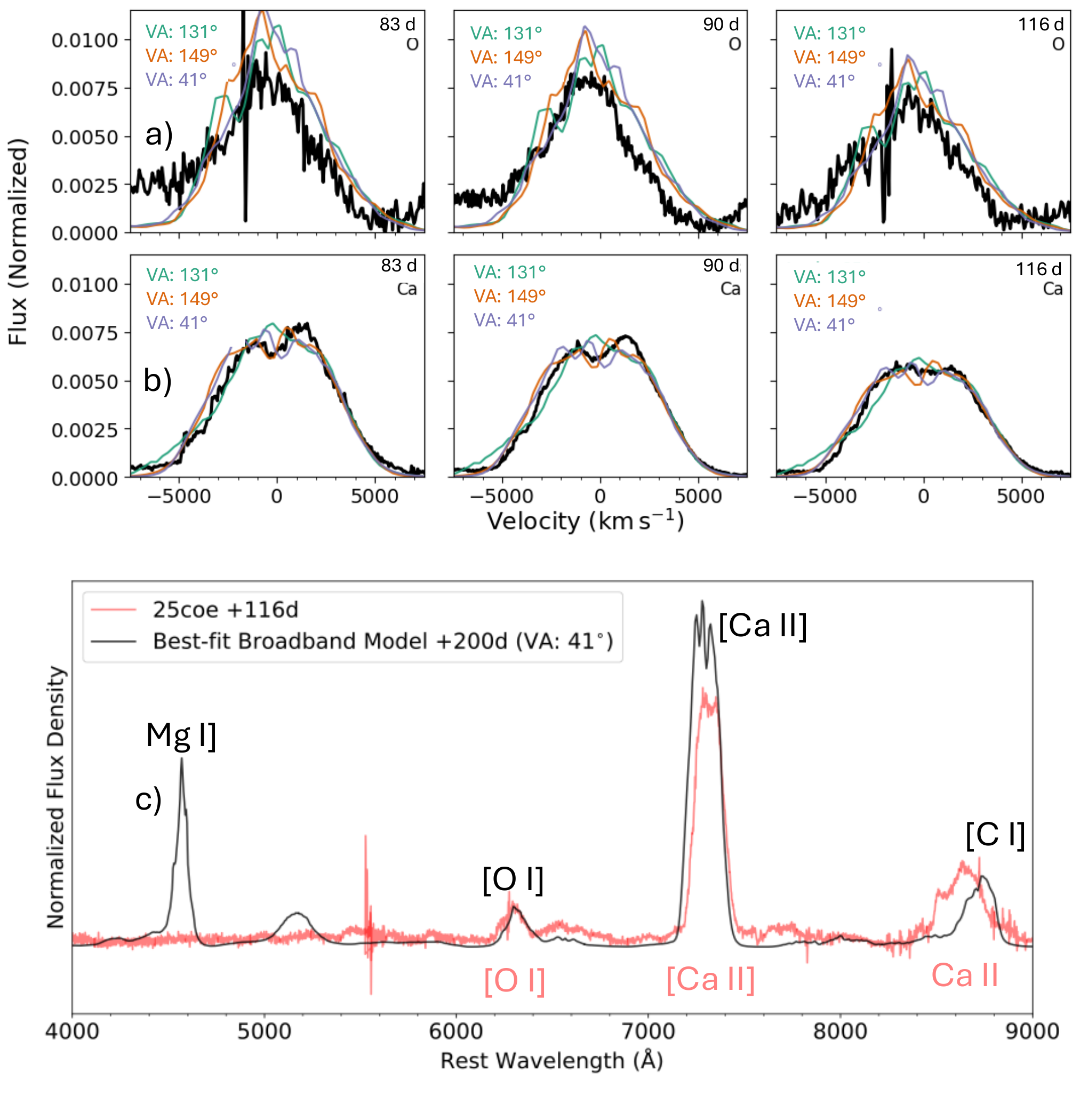}
    \caption{Best viewing angles for a simultaneous fit of the synthetic spectra from the HEC-33L explosion model with the observed [\OI] (\textit{panel a}) and [\CaII] (\textit{panel b}) doublet profiles (in black) at days 83, 91, and 116 after explosion. The $\theta$ angle measured from the north pole corresponding to each model is marked. The velocities are centered around $\lambda$6316 for [\OI] and around $\lambda$7304 for [\CaII]. \textit{Panel c}: Comparison between normalized broadband synthetic spectrum ($\theta$ = 41$^{\circ}$) and the latest observed nebular spectrum.}
    \label{fig:nebular_modeling}
\end{figure*}

The double-peaked [\CaII] line profile in the nebular spectra of SN\,2025coe (Figure \ref{fig:nebular}) is rarely observed among CaSTs ($\lesssim$8\,\%) and the broader SN\,Ib/c class  \citep[e.g.,][]{Milisavljevic10, Milisavljevic17, Roy13, Modjaz14, Jacobson-Galan22}. Among CaSTs, in iPTF15eqv, the double component [\CaII] was explained as a potential consequence of the observer's line of sight \citep{Milisavljevic17}. Multipeaked [\OI] in SNe\,Ib/c have been interpreted to represent ejecta asymmetry \citep[e.g.,][]{Maeda08, Modjaz08, Taubenberger09}, although in many of these cases the multiple peaks have been noted to be actually due to the doublet nature of [\OI] in conjunction with ejecta clumping \citep{Milisavljevic10}. In SN\,2025coe, the continuum-subtracted renormalized line profiles of [\CaII] and [\OI] at 59 days after explosion in velocity space shows that some Ca and O ejecta might be co-located while they are also likely in distinct locations (Figure \ref{fig:OI_CaII_profile}). The red-component of [\CaII] cannot be explained by the $\lambda$7291 and $\lambda$7323 lines and is thus likely a geometric representation of Ca ejecta distribution and/or viewing angle effects. 

Motivated by these nebular diagnostic studies and observations of SN\,2025coe, in this section we simultaneously compare our observed line profiles of [\CaII] and [\OI] with synthetic line profiles at different viewing angles from a low-energy asymmetric three-dimensional (3D) explosion model to match the observed explosion characteristics of SN\,2025coe. For this purpose we employed the 3D non-local thermodynamic equilibrium (NLTE) spectral synthesis code \texttt{ExTraSS} (EXplosive TRAnsient Spectral Simulator; see \citealt{vanBaal25code} for a full code description) in the optically thin limit. We used the synthetic nebular-phase spectra from one of the He-core progenitor models, which are described by \citet{vanBaal23,vanBaal24}.

Based on explosion parameters as discussed in Section \ref{sec:7.1}, we use the low-energy $3.3\,M_\odot$ He-core progenitor, HEC-33L \citep{vanBaal24}. This model entails the asymmetric explosion of the He-core progenitor with an  energy of $4.7\times10^{50}$\,erg and a total ejecta mass of $1.204\,M_\odot$ (of which $0.045\,M_\odot$ is $^{56}$Ni), reasonably close to the estimated parameters for SN\,2025coe (see Sections \ref{sec:5.2} and \ref{sec:7.1}; Table \ref{tab:bolometric_fitting}). The explosion was performed with the \texttt{Prometheus-HotB} CCSN code \citep{Fryxell91,Muller91a,Muller91b,Janka96,Kifonidis03,Kifonidis06,Scheck06,Arcones07,Wongwathanarat10,Muller12,Wongwathanarat13,Wongwathanarat15, Wongwathanarat17}
with the hydrodynamical simulation continued for 1001\,s post-bounce.

Assuming homologous expansion, we mapped the output ejecta to \texttt{ExTraSS} and fast-forwarded to the nebular phase at 200 days post-explosion. With \texttt{ExTraSS} we first determined the radioactive decay of $^{56}$Ni and inferred the nonthermal electron distribution using the Spencer-Fano method developed by \citet{Kozma92}.
The model accounts for thermal and nonthermal collisions, radiative decay (in the Sobolev approximation), and recombination. It also accounts for local re-ionization by the trapped UV photons, but excludes radiation transport between regions. The spectra were then computed from the NLTE level populations under the globally optically thin limit.

SN\,2025coe is estimated to have lower ejecta mass than HEC-33L and thus will have lower densities, lessening the impact of the timing difference to HEC-33L as the model spectra were computed at 200 days. More details on the explosion of HEC-33L, the ejecta profiles and composition, and the explosion simulation as well as the nebular phase modeling are presented by \citet{vanBaal24}. 

For a consistent picture of the explosion scenario, we fit both line profiles ([\OI] $\lambda\lambda$6300, 6364 and [\CaII] $\lambda\lambda$7291, 7323) at the same time to find the viewing angles that can explain the observations simultaneously using a simple $\chi^2$ minimization. From the model spectra, the region within $\pm\,7500\,\text{km}\,\text{s}^{-1}$ of the line center was selected, and the data in this region were interpolated to the same resolution as the observations. The flux in this region was normalized for both the model and the observations. Then, the $\chi^2$ for every viewing angle (in a $20\times20$ grid in the azimuthal and polar coordinates) was calculated for both line profiles separately, and added together to find the angles which had the best overall fit. At days 83 and 116 after explosion, the [\OI] profile has some sky-subtraction artifacts still present in the data, but this does not impact the fitting in a significant manner.

In Figure \ref{fig:nebular_modeling}, the three best-fit viewing angles for each epoch are shown together with the observed profiles (in black). Three viewing-angle models, (1) $\theta$ = 131$^{\circ}$, $\phi$ = 171$^{\circ}$; (2) $\theta$ = 149$^{\circ}$, $\phi$ = 333$^{\circ}$; and (3) $\theta$ = 41$^{\circ}$, $\phi$ =  153$^{\circ}$ (where $\theta$ is the polar angle measured from the north pole and $\phi$ is the azimuthal angle of rotation around the $z$ axis), provide statistically equivalent good fits to both [\CaII] and [\OI] across all three observed epochs (Figure \ref{fig:nebular_modeling}).  Since [\CaII] is much stronger than [\OI], the $\chi^{2}$ values are weighted in our comparison. We find that the model spectrum with a viewing angle $\theta$ = 149$^{\circ}$, $\phi$ = 333$^{\circ}$ (orange in Figure \ref{fig:nebular_modeling}) provides the closest match to the observed [\CaII] profile at $v$ = 0 on days 83 and 91 (after explosion) when the two peaks are clearly identified. By day 116 (after explosion) the [\CaII] becomes more flat-topped. The He-core explosion models from \citet{vanBaal23,vanBaal24} have 3D ejecta distributions; as such, finding preferential distributions of ejecta in SN\,2025coe along particular viewing angles gives a strong indication of ejecta asymmetries in the observed SN.

Based on these results, we propose that both the single-component [\OI] doublet and double-component [\CaII] doublet in the nebular spectra of SN\,2025coe can be explained through asymmetries in the ejecta distribution.   From the model grid, we estimate $\sim$10\% of the synthesized spectra to have a double component for the [\CaII] doublet, consistent with the empirical expectation of $\sim$8\% from CaST observations. One caveat to note here is that while the optical line profile changes shape over time, the changes in model profiles are not as pronounced. Future nebular samples of CaSTs will be crucial in further constraining model parameters for accurate description of the ejecta geometry. Nevertheless, the reasonably good fit of the models (based on the asymmetric explosion of a low-mass He star) with the observed asymmetric line profiles in conjunction with a low observed ejecta mass could be indicative of SN\,2025coe's progenitor being a low-mass massive star that exploded asymmetrically. 

In the bottom panel of Figure \ref{fig:nebular_modeling}, we show one of the best-fitting synthetic nebular spectra compared to the observed broadband nebular spectrum at day 116. While the [\CaII] and [\OI] line profiles are well explained with our simulated spectra, there are still a few caveats. Recent literature on simulations of exploding low-mass He-core progenitors have predicted the presence of strong [N~II] doublet emission in the nebular spectra \citep{Dessart23, Barmentloo24} which has not yet been observed in any CaST. Our synthetic models do not account for [N~II] and thus adds a caveat in associating the observed low ejecta mass of SN\,2025coe with a low-mass He-core. Additionally, the presence of Mg~I] $\lambda$4571 expected in the nebular spectra of SESNe \citep{Jerkstrand15} and also predicted in our synthetic spectrum is not detected in the latest nebular spectrum of SN\,2025coe. While Mg~I] has been observed in the nebular spectra of SESNe and acts as a ubiquitous cooling line \citep{Jerkstrand17}, in CaSTs their strength could be suppressed if majority of the cooling occurs through the forbidden [\CaII] emission instead \citep{Polin21}. Also, for low-mass He-core progenitors, contribution of Mg~I] to the overall flux is predicted to be only significant at much later epochs \citep[$\sim$400\,d;][]{Dessart23} compared to the latest spectrum of SN\,2025coe discussed in this work at 116 days after explosion. Thus, nondetection of Mg~I] in SN\,2025coe might not be a challenge for a low-mass He-core progenitor. The bump in the synthetic model around 5200\,\AA\ comes from Fe line transitions (both neutral and singly ionized) that were not observed in SN\,2025coe. As there is a time delay between the data (day 116) and the model (day 200), it is also not unexpected that the Ca~II triplet is significantly weaker in the model and [C~I] $\lambda$8727 is the more dominant component in the model at these wavelengths.

Recently, a progenitor model involving a binary WD system being embedded in an environment polluted by recurrent helium novae was invoked to explain observations of the CaST SN\,2023xwi \citep{Touchard-Paxton25}. In this AM Canum Venaticorum (AM CVn)-like system of binary WDs, a CO-WD accretes He from its companion. Periodically the system undergoes a He nova, causing ejection of a He layer and polluting the environment of the binary system. A consequence of such recurrent He novae in the model system is the distribution of O and Ca ejecta into low-velocity central and high-velocity polar outflow-like emitting regions, where the observed strength of the latter emission will be viewing-angle dependent. Thus, alternatively, the transition observed in the double component of [\CaII] and dependence on viewing angles (Figure \ref{fig:nebular_modeling}) could be a consequence of tracing these two different velocity regions. Future model nebular spectra assuming an AM~CVn-like progenitor would be a key test of this possible channel. These results suggest that probing the distribution of ejecta through nebular spectroscopy is perhaps an important consideration in understanding the progenitor-channel diversity for CaSTs.

\begin{figure*}
    \centering
    \includegraphics[width=\textwidth]{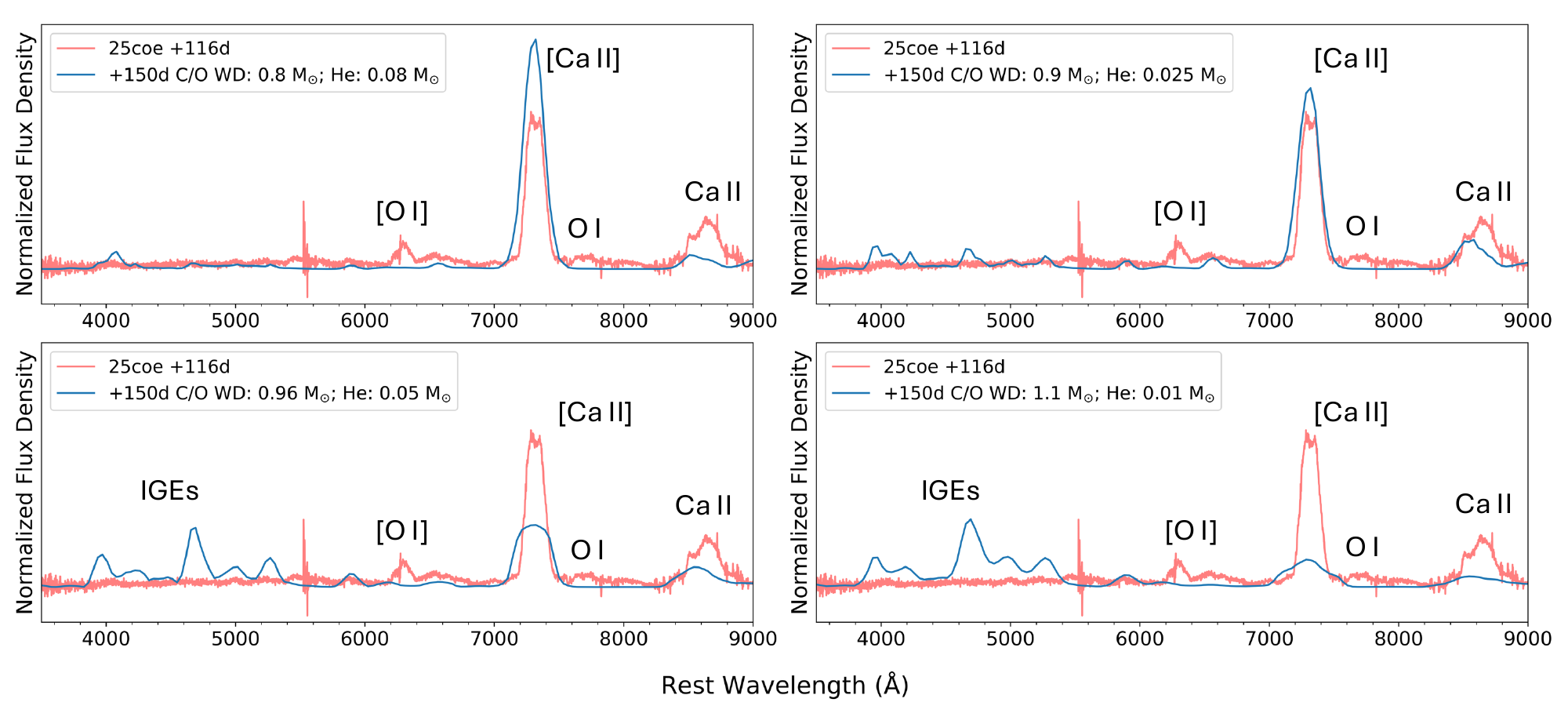}
    \caption{Comparison of the latest nebular spectrum of SN\,2025coe at day 116 (after explosion) compared with nebular SN\,Ia double-detonation models of sub-Chandrasekhar-mass CO WDs with He shells from \cite{Polin21}. The observed SN\,2025coe spectrum lacks Fe-group element lines that are prominent for the higher mass WDs models. A WD of 0.9\,\Mdot{} with a He shell of 0.025\,\Mdot{} provides the closest match to our data.}
    \label{fig:thermonuclear_comparison_Polin}
\end{figure*}

\begin{figure*}
    \centering
    \includegraphics[width=\textwidth]{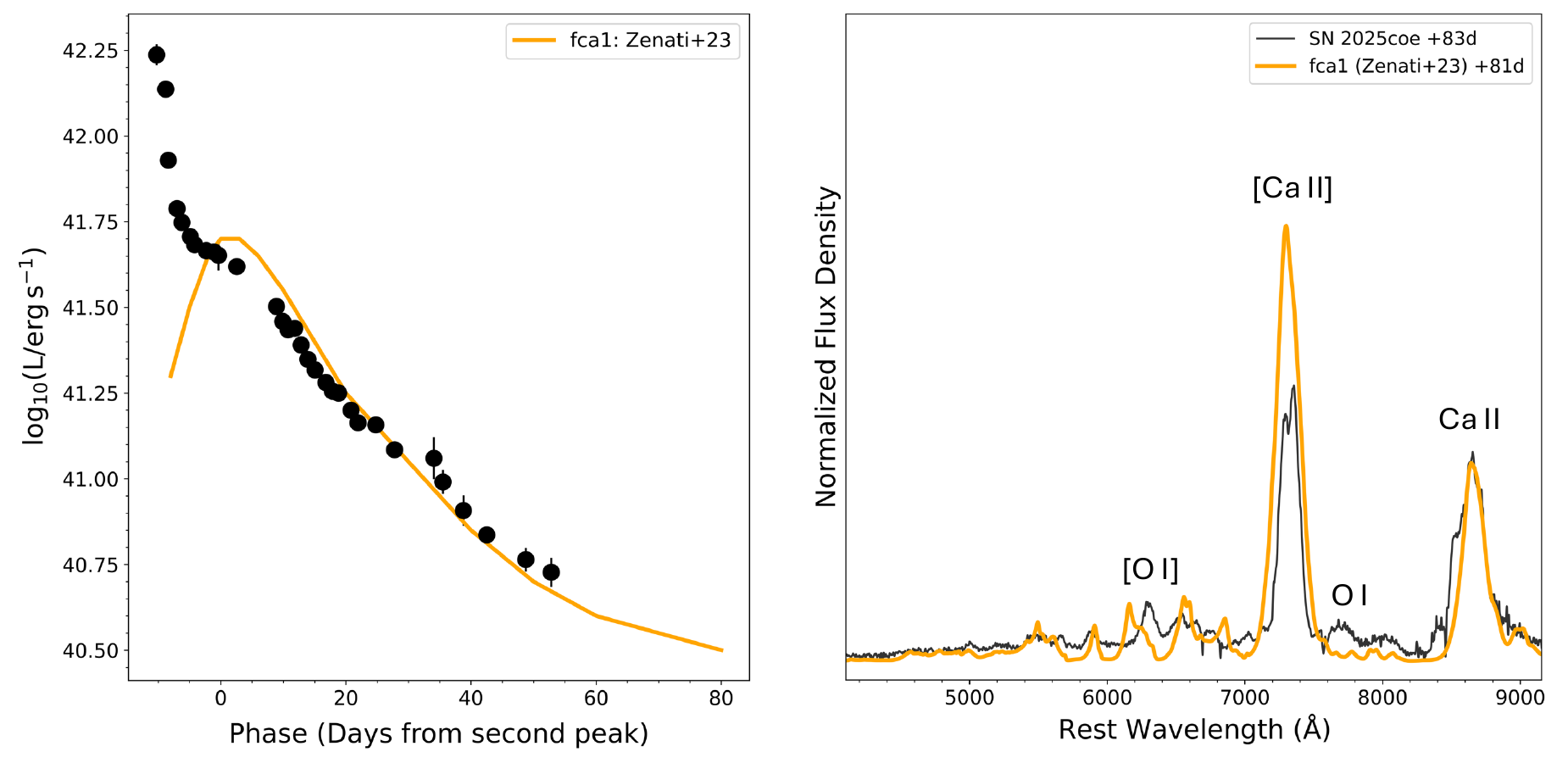}
    \caption{\textit{Left}: Comparison between the bolometric light curve of SN\,2025coe and the predicted luminosity from the disruption of a C/O WD by a hybrid He-C/O WD model (fca$_{1}$) as described by \cite{Zenati23}. The model explains the second peak and nebular luminosities well. \textit{Right}: Comparison between a nebular spectrum of SN\,2025coe and a synthetic spectrum from the fca$_{1}$ explosion model at comparable epochs (after explosion). Several emission lines are reproduced in the model; however, it overestimates the strength of [\CaII] and underestimates the strength of all O emission lines.}
    \label{fig:thermonuclear_comparison_Zenati}
\end{figure*}

\subsection{Progenitor Channels} \label{sec:7.2}
SN\,2025coe falls firmly within the definition of the CaST class based on our photometric and spectroscopic analyses. In this section we discuss a few potential progenitor scenarios that can plausibly explain these observations. As with other CaSTs, broadly these fall into two categories: (a) core collapse of a low-mass massive star, and (b) thermonuclear explosion of a WD in a binary system.

\subsubsection{Low-Mass Massive Star} \label{sec:7.2.1}
Modeling the bolometric light curve of SN\,2025coe suggests that like other double-peaked CaSTs, there are two power sources: (1) early emission from the cooling of a compact shocked envelope ($\sim$6--40\,$R_{\odot}$) and/or compact CSM ($R \approx 10^{15}$ cm) around the progenitor, and (2) radioactive decay of $^{56}$Ni and $^{56}$Co. 

One plausible physical scenario to explain these observations is the gravitational collapse of a low-mass massive star having a compact envelope around it. Based on results in Section \ref{sec:7.1}, we find that a total $M_{\mathrm{ej}}$ of $\sim$ 0.4--0.5\,$M_{\odot}$ and a compact envelope of 0.1--0.2\,$M_{\odot}$ can explain the light curve of SN\,2025coe. Stellar evolution models have predicted that He stars with pre-SN core masses within $\sim$2.5\, $M_{\odot}$, which had all of their H-rich envelope stripped off through a binary companion, can be mapped to the lower zero-age main-sequence (ZAMS) mass end of massive stars \citep{Woosley19, Laplace20}.   

In a core-collapse scenario, the resulting remnant neutron stars typically have masses of 1.3--1.7\,\Mdot{} \citep[e.g.,][]{Antoniadis16}. Combining this with the ejecta and envelope mass estimates for SN\,2025coe, the pre-SN mass would be $\sim$1.8--2.4\,\Mdot{}. This suggests that the evidence of an early shock-cooling envelope along with estimated explosion properties can be potentially the result of the Fe CCSN of a low-mass massive star (8\,$M_{\odot} < M_{\mathrm{ZAMS}} < 12\,M_{\odot}$). 

An alternative massive progenitor scenario is the possibility that SN\,2025coe comes from an electron-capture SN. However, the estimated $^{56}$Ni mass of $\sim$0.014 \Mdot{} from our bolometric fitting (Section \ref{sec:7.1}) is an order of magnitude higher than what is expected from such a scenario \citep[$\sim$10$^{-3}$\,\Mdot{};][]{Moriya14}, so we disfavor this progenitor channel for SN\,2025coe.

Lack of any H signature at early times suggests that the CSM/envelope around SN\,2025coe is likely H-free (or at least significantly H-poor); thus, another candidate could be a He-star binary system capable of producing a SN\,Ib-like explosion \citep[see, e.g.,][]{Yoon17, Jung22}. Lack of any high-ionization features from interaction in the optical spectra (Figure \ref{fig:opt_spectra}) could be a sign of an asymmetric or clumpy distribution of CSM such that narrow features from the interaction are not observed from several viewing angles \citep{Smith17}. 

The sustained X-ray observations and larger inferred mass of CSM \citep[$\sim$0.1\,\Mdot{};][]{Kumar26} compared to other X-ray detections of SN\,2019ehk \citep{Jacobson-Galan20b} and SN\,2021gno \citep{Jacobson-Galan22} indicates significant mass loss either through binary mass transfer or mass eruptions before explosion, which is more likely to be a massive-star attribute than a typical WD progenitor. 

 That said, a significantly large offset from the potential host galaxy NGC 3277 coupled with a low SFR at the explosion site (see Section \ref{sec:4}) is a challenge to the massive-star interpretation. Deep observations of the explosion sites of several CaSTs have previously ruled out surviving massive-star binary companions, underlying dwarf galaxies, and globular clusters below the detection limit, suggesting one channel of progenitors to be high-velocity kicked systems from older stellar populations \citep{Lyman14}. Alternatively, stellar populations in highly extended haloes of early-type galaxies have also been proposed to explain the offsets of CaSTs \citep{PeretsBenamini21}.  The potential host galaxy of SN\,2025coe, NGC 3277, is an early-type spiral with an isophotal radius ($R_{25}$) of $\sim 0.98'$  \citep[i.e., $\sim$7.1 kpc at $D \approx 25$\,Mpc;][]{deVaucouleurs91}. Thus, depending on where the progenitor star system was born in NGC 3277, the total distance to be covered to the site of SN\,2025coe is $\sim 27$--34 kpc. 
 
 The typical lifetime of a massive-star progenitor before exploding is $\sim$10 Myr \citep[e.g.,][]{Schaller92}; to reach an offset of $\sim$27--34 kpc, it would have needed a velocity of $\sim$2600--3300 km s$^{-1}$, unrealistically large for even hypervelocity stars at 500--1000 km s$^{-1}$ \citep[e.g.,][]{Hills88, Brown15}. However, as discussed in Section \ref{sec:4}, we cannot fully exclude the possibility of the progenitor originating in one of the six nearby extended faint sources at offsets $\lesssim$3 kpc. Correspondingly, to have traveled to the explosion site from the closest of these sources at a projected offset of $\sim$0.8 kpc, the massive progenitor star would need to have traveled at a velocity of $\sim$80 km s$^{-1}$, far more typical  of runaway stars \citep[e.g.,][]{Hoogerwerf01, Eldridge11, Renzo19}. It is possible that the progenitor of SN\,2025coe was a runaway star ejected from its site of birth through dynamical evolution in a binary system where the primary star exploded \citep{Eldridge11}. Thus, a large velocity need not be necessary to explain where SN\,2025coe exploded if the birth site of its progenitor was in a satellite galaxy around NGC 3277.

\subsubsection{Thermonuclear White Dwarf Explosion} \label{sec:7.2.2}

The low ejecta mass and large offset of SN\,2025coe from its potential host NGC 3277 is similar to many other objects of the CaST class, some of which have been suggested to originate from the thermonuclear disruption of a WD \citep[e.g.,][]{Perets10, Kasliwal12, Sell15, Sell18, De20, Jacobson-Galan20b, Jacobson-Galan22}. Low-mass ($\lesssim$0.98\,$M_{\odot}$) double-detonation models with a small mass fraction of Ca (at $\sim$1\,\%) have produced nebular spectra that cool primarily through [\CaII] emission, making them a viable progenitor channel for CaSTs \citep{Polin21}. We compare these models at day 150 with the latest nebular spectrum of SN\,2025coe at day 116 after explosion (Figure \ref{fig:thermonuclear_comparison_Polin}). All model spectra were normalized by the integrated flux within the wavelength range of the observed spectrum. We find that a WD mass of 0.9\,$M_{\odot}$ with a surrounding He shell of 0.025\,$M_{\odot}$ matches the observed intensity of [\CaII] most closely. However, these models do not predict the [\OI] and O\,I emission lines that are clearly observed in SN\,2025coe. The best-matching model also overpredicts the strength of IGEs in the spectrum compared to observations. This suggests all high-mass WDs that produce significant IGEs after a thermonuclear explosion can be safely excluded as progentiors of SN\,2025coe. 

Oxygen emission lines have been observed in the nebular spectra of several WD explosions believed to be triggered in a violent merger scenario \citep[e.g.,][]{Taubenberger13, Kromer2013, Mazzali22, Dimitriadis23, Siebert24, Kwok2024}. Tidal disruption of a hybrid He-C/O WD by a normal C/O-WD or another hybrid He-C/O-WD through violent mergers can induce a He detonation that can lead to a CaST-like event \citep{Bobrick17, Perets19, Zenati19a, Zenati19b, Zenati23}. Such double-degenerate hybrid scenarios have been previously invoked to explain the observed features of SN\,2021gno and SN\,2021inl \citep{Jacobson-Galan22} as well as SN\,2019ehk \citep{Jacobson-Galan20b}. 

To explore this possibility, we compare the bolometric light curve and the observed nebular spectra of SN\,2025coe with the corresponding light curve and synthetic spectrum predicted from the disruption of a C/O WD by a hybrid He-C/O WD model (fca$_{1}$) as described by \cite{Zenati23} (Figure \ref{fig:thermonuclear_comparison_Zenati}). While the early excess due to shock cooling or CSM interaction is not explored by fca$_{1}$, the second peak and nebular luminosities are well matched with this model (Figure \ref{fig:thermonuclear_comparison_Zenati}; left panel). 

Spectroscopically, we compare the fca$_{1}$ model spectra of \cite{Zenati23} with the observations by normalizing the model spectra across the observed broadband wavelengths. The model  reproduces many of the observed lines in SN\,2025coe; however, the primary difference is the significantly weaker [\OI] and O\,I lines (Figure \ref{fig:thermonuclear_comparison_Zenati}; right panel). The offset in abundance between Ca and O can be large in different regions of the ejecta when the material is not well mixed. Thus, weaker O emission in the models could be due to artificial mixing of Ca and O ejecta introduced when mapping 2D simulation results into a 1D NLTE code used to model the spectra resulting from radiative-transfer analysis \citep{Zenati23}. The strengths of IGEs, He, and [\CaII] emission lines are overestimated. The single component of [\CaII] doublet in the model vs. the observed two-component [\CaII] doublet is likely due to the asymmetry in the ejecta distribution which is not accounted for in the models. Nevertheless, the synthetic spectra can reproduce many of the observed lines in general. Other NLTE spectral models for low-mass C/O and He WD mergers have also predicted observed features, including strong [\CaII] and weak [\OI] features at nebular phases \citep{Callan25}. Their synthesized spectra overpredict the strength of the optical He~I features, and suggest a significant contribution from Ti~II (which is not observed in SN\,2025coe), resulting in a substantially redder SED than most CaSTs at peak. 

Prior to disruption of a hybrid He-C/O WD, mass transfer from the secondary to the primary can form an accretion disk owing to its sufficiently large angular momentum and may produce continuous outflow in the form of disk winds \citep[][]{Zenati19a}.  This outflow can expand outward to form CSM \citep{Raskin_Kasen13}, a fraction of which might accumulate around the primary WD, forming a low-mass envelope \citep{Shen12, Schwab16}. While such mass transfer could potentially explain the presence of material in the local environment of SN\,2025coe, the predicted mass to be accumulated from this scenario is significantly lower than what we inferred from the shock-cooling envelope (Section \ref{sec:7.1}). The independently estimated CSM mass from longer sustained X-ray detections of SN\,2025coe than SN\,2021gno or SN\,2021inl \citep{Kumar26} is also too high to be accounted by such mass transfer.

An alternate possibility to justify the CSM quantity in the thermonuclear scenario could be a recurrent He nova AM-CVn system \citep{Touchard-Paxton25} polluting the environment of a binary WD system. Considering this progenitor system seems to offer an explanation for both the presence of close-in CSM around the progenitor system and the observed velocity distribution of [\CaII], it is a strong contender to understand the origin of CaSTs. 

While each model explains parts of the observations, no single model progenitor scenario can explain all the observed properties of SN\,2025coe. Though the large offset from the potential host galaxy favors an older progenitor undergoing a thermonuclear explosion, the presence of significant CSM and ejecta asymmetry favor the explosion of a low-mass massive star. 

\section{Summary and Conclusions} \label{sec:8}

Our multiwavelength and extensive study of SN\,2025coe in this paper adds additional constraints on the progenitor channels that can produce CaSTs. We characterize the explosion parameters, an early blue excess, strong [\CaII] and weak [\OI] emission lines, and line profile asymmetries in the nebular spectra to establish SN\,2025coe's place among the growing sample of CaSTs. Here we summarize our results. 

\begin{enumerate}
    \item SN\,2025coe is the second-closest CaST observed at $D \approx 25$ Mpc with a significant projected offset of $\sim$34 kpc from its potential early-type spiral host galaxy, NGC 3277. However, we cannot rule out the actual host galaxy being a satellite galaxy of NGC 3277 owing to the presence of $\sim$6 nearby faint extended sources at offsets $\lesssim$3 kpc. 
    \item Detected within $\sim$0.6 days after explosion, the optical light curve of SN\,2025coe shows a double-peaked structure, a characteristic feature among some CaSTs with high-cadence early observations.  The early blue excess in the bolometric light curve can be reproduced either by shock-cooling emission from $\sim$0.1\,\Mdot{} of a compact envelope ($R_\mathrm{env}$\,$\sim$6--40\,$R_{\odot}$) or by shock interaction with close-in CSM ($R_{\mathrm{CSM}} \lesssim 6\,\times 10^{14}$\,cm). The second peak, powered by radioactive decay, occurs at $\sim$11 days after explosion and indicates a low-luminosity ($M^{\mathrm{peak}}_{o} \gtrsim 15.5$\,mag). Bolometric modeling of this peak indicates a low ejecta mass ($M_{\mathrm{ej}} \approx 0.4$--0.5\,$M_{\odot}$) and small amounts of synthesized $^{56}$Ni ($M_{^{56}\mathrm{Ni}} \approx 1.4 \times 10^{-2}$\,\Mdot{}) in SN\,2025coe. The low luminosity, fast decline, and low ejecta and $^{56}$Ni masses are consistent with observed properties of other CaSTs.
    \item SN\,2025coe undergoes rapid spectral evolution from the photospheric phase dominated by He~I P-Cygni lines to a nebular phase marked by strong [\CaII] and weak [\OI] ([\CaII]/[\OI] $\gtrsim 10$). Abundance estimates of O and Ca from the latest nebular spectrum emphasizes that these explosions are not ``rich" in Ca compared to O; rather, they have strong Ca emission likely arising from the degree of mixing, ionization, and excitation conditions in the ejecta.
    \item We report the development of an asymmetric line profile in the nebular phase, specifically the double-component [\CaII] doublet contrasted with the single-component [\OI] doublet. The similarities and differences between line shapes of [\OI] and [\CaII] could be an indication of a mix of colocated and distinct distributions of Ca and O ejecta in an asymmetric explosion.
    \item Using nebular line synthesis from hydrodynamical modeling of a core-collapse scenario, we simultaneously fit the [\OI] and [\CaII] line profiles across several nebular epochs. We find that the asymmetric core collapse of a low-mass He star ($\sim$3.3\,\Mdot{}) with viewing-angle dependence best explains the observed line profiles.
    \item No current single progenitor channel can explain all observed features of SN\,2025coe. If NGC 3277 is the host, the large projected offset favors a thermonuclear origin. Violent merger scenarios involving hybrid He-C/O WD systems can also potentially explain the observed luminosity and spectral features, although they often overpredict the presence of IGEs and underpredict the strength of O emission lines. The presence of $\sim$0.1\,\Mdot{} of CSM cannot be naturally explained by these scenarios, although pollution from a recurring helium nova contributing to the environment of an exploding WD cannot be entirely ruled out.
    \item On the other hand, the core collapse of a low-mass massive star in a binary system could be a more natural explanation for the presence of $\sim$0.1\,\Mdot{} inside a compact envelope or close-in CSM, and observed ejecta asymmetry. However, the lack of a significant SFR, together with poorly constrained redshifts of the faint extended sources near the site of SN\,2025coe, argue against this scenario.
\end{enumerate}

We find that no current single progenitor channel model explains all the observations of SN\,2025coe. The early interaction and ejecta asymmetries put constraints on potential progenitor models. Future nebular spectral modeling across both thermonuclear and core-collapse scenarios will need to account for these constraints to pin down the progenitor channel diversity suggested by observations of CaSTs. Like other members of the double-peaked light curve CaSTs, SN\,2025coe underscores the need for multiwavelength follow-up observations at both early and late times to understand the diversity of the overall CaST landscape.

\begin{acknowledgments}

\end{acknowledgments}

We thank the anonymous referee for offering a careful and constructive consideration of our work.
This research has made use of the \cite{NED},
which is operated by the Jet Propulsion Laboratory, California Institute of Technology, under contract with the
National Aeronautics and Space Administration (NASA). We thank Dr. Yossef Zenati for sharing comparison models for our discussion.

S.V. and the UC Davis time-domain research team acknowledge support from National Science Foundation (NSF) grant AST-2407565. M.M. the METAL group at UVA  acknowledges support in part from ADAP program grant 80NSSC22K0486, from  NSF grant AST-2206657, from NASA/{\it HST} program GO-16656, and from the NSF under Cooperative Agreement 2421782 and the Simons Foundation grant MPS-AI-00010515 awarded to the NSF-Simons AI Institute for Cosmic Origins -- CosmicAI, https://www.cosmicai.org/. R.B.W. is supported by the NSF Graduate Research Fellowship Program under grant 2234693 and by the Virginia Space Grant Consortium.  J.E.A. is supported by the international Gemini Observatory, a
program of NSF's NOIRLab, which is managed by the Association of Universities for Research in Astronomy (AURA) under a cooperative agreement with the NSF, on behalf of the Gemini partnership of Argentina, Brazil, Canada, Chile, the Republic of Korea, and the United States of America. K.A.B. is supported by an LSST-DA Catalyst Fellowship; this publication was thus made possible through the support of grant 62192 from the John Templeton Foundation to LSST-DA.
Time-domain research by the University of Arizona team and D.J.S. is supported by NSF grants 2108032, 2308181, 2407566, and 2432036, as well as by the Heising-Simons Foundation under grant 20201864. 
The research group of A.V.F. at UC Berkeley acknowledges financial assistance from  Gary and Cynthia Bengier, Clark and Sharon Winslow, Alan Eustace and Kathy Kwan (W.Z. is a Bengier-Winslow-Eustace Specialist in Astronomy), Timothy and Melissa Draper, Briggs and Kathleen Wood, Ellyn and Alan Seelenfreund (T.G.B. is Draper-Wood-Seelenfreund Specialist in Astronomy), and numerous other donors. A.J. and B.vB acknowledge support from the European Research Council (ERC) under the European Union’s Horizon 2020 Research and
Innovation Programme (ERC Starting grant 803189, PI A. Jerkstrand).

This work makes use of data from the Las Cumbres Observatory global telescope network, which is supported by NSF grants AST-1911225 and AST-1911151. TNOT was sponsored by the Natural Science Foundation of Xinjiang Uygur Autonomous Region under grant 2024D01D32, Tianshan Talent Training Program grant 2023TSYCLJ0053，and the National Natural Science Foundation of China NSFC grant 12373038. A major upgrade of the Kast spectrograph on the Shane 3\,m telescope at Lick Observatory, led by Brad Holden, was made possible through gifts from the Heising-Simons Foundation,  
William and Marina Kast, and the University of California Observatories.
Research at Lick Observatory is partially supported by a gift from Google.
Some of the data presented herein were obtained at Keck Observatory, which is a private 501(c)3 nonprofit organization operated as a scientific partnership among the California Institute of Technology, the University of California, and NASA. The Observatory was made possible by the generous financial support of the W. M. Keck Foundation. 
The authors wish to recognize and acknowledge the very significant cultural role and reverence that the summit of Maunakea has always had within the Native Hawaiian community. We are most fortunate to have the opportunity to conduct observations from this mountain.              We appreciate the expert   
assistance of the staff at 
the various observatories where data were obtained.     

\appendix

\section{Spectra Log} \label{sec:Appendix A}

\startlongtable
\begin{deluxetable}{cccccc}
\tabletypesize{\footnotesize}
\tablenum{A1}
\tablecaption{Optical Spectra of SN\,2025coe\label{tab:optical_spectralog}}
\tablewidth{0pt}
\tablehead{
\colhead{UTC Date \& Time (hh:mm:ss)} & \colhead{Modified Julian Date (days)} &
\colhead{Phase (days)$^{\star}$} &
\colhead{Telescope} & \colhead{Instrument} & \colhead{Wavelength Coverage (\AA)}}

\startdata
2025-02-25 11:57:12 & 60731.50 & 1.20 & FTN & FLOYDS & 3400 -- 10000 \\
2025-02-26 07:24:58 & 60732.31 & 2.01 & Lick/Shane & Kast & 3600 -- 10700\\
2025-02-28 11:04:40 & 60734.46 & 4.16 & FTN & FLOYDS & 3400 -- 10000\\
2025-03-01 13:44:06 & 60735.57 & 5.27 & FTN & FLOYDS & 3400 -- 10000 \\
2025-03-03 12:19:55 & 60737.51 & 7.21 & FTN & FLOYDS & 3400 -- 9000 \\
2025-03-04 11:16:48 & 60738.47 & 8.17 & Lick/Shane & Kast & 3600 -- 10700\\
2025-03-04 18:28:21 & 60738.77 & 8.47 & Xinglong  & BFOSC & 3780 -- 8910 \\
2025-03-06 09:45:08 & 60740.41 & 10.11 & FTN & FLOYDS & 3400 -- 9000 \\
2025-03-08 11:20:29 & 60742.47 & 12.17 & Xinglong & BFOSC & 3780 -- 8920 \\
2025-03-09 08:06:43 & 60743.34 & 13.04 & Lick/Shane & Kast & 3600 -- 10700\\
2025-03-17 06:50:11 & 60751.28 & 20.98 & FTN & FLOYDS & 3400 -- 9400 \\
2025-03-22 05:49:55 & 60756.24 & 25.94 & Lick/Shane & Kast & 3600 -- 5600\\
2025-03-23 11:27:50 & 60757.48 & 27.18 & FTN & FLOYDS & 3400 -- 9400 \\
2025-03-28 07:05:20 & 60762.30 & 32.0 & FTN & FLOYDS & 3400 -- 9400 \\
2025-03-29 05:53:35 & 60763.24 & 32.94 & MMT & Binospec & 4120 -- 9200 \\
2025-04-03 06:22:29 & 60768.27 & 37.97 & FTN &FLOYDS & 3400 -- 9000 \\
2025-04-06 05:44:49 & 60771.24 & 40.94 & Bok & Boller \& Chivens & 4000 -- 8000 \\ 
2025-04-09 06:38:46 & 60774.28 & 43.98 & Lick/Shane & Kast & 3300 -- 10900 \\
2025-04-22 07:02:25 & 60787.29 & 56.99 & Lick/Shane & Kast & 3300 -- 10900 \\
2025-04-23 00:34:13 & 60788.02 & 57.72 & SOAR & Goodman-RED & 4930 - 8900\\
2025-04-24 08:07:11 & 60789.33 & 59.03 & Keck-I & LRIS & 3130 -- 10300 \\
2025-04-26 05:25:13 & 60791.22 & 60.92 & MMT & Binospec & 4120 -- 9200\\
2025-05-18 03:54:42 & 60813.16 & 82.86 & MMT & Binospec & 4120 -- 9200\\
2025-05-25 08:37:30 & 60820.36 & 90.06 & Keck-I & LRIS & 3130 -- 10250 \\
2025-06-20 06:26:43 & 60846.27 & 115.97 & Keck-II & KCWI & 3380 -- 9400 \\
\enddata
\tablenotetext{\star}{Phase from explosion}
\end{deluxetable}

\section{Bolometric Fitting} \label{sec:Appendix B}

\captionsetup[table]{width=0.8\textwidth}


\begin{table*}[!b]
    \centering
    \begin{tabular}{l|c|c|c}
    \hline
    \hline
    \textbf{Parameter}   & \textbf{Piro+21 \& Radioactivity} & \textbf{Margalit+22 \& Radioactivity}  & \textbf{Units} \\
    \hline
    Envelope Radius ($R_\mathrm{env}$)   & $21.56^{+15.81}_{-8.62}$ & $6.61^{+0.27}_{-0.26}$ &  $R_{\odot}$ \\
    Envelope Mass ($M_\mathrm{env}$)        & $0.18^{+0.05}_{-0.04}$ & $0.17^{+0.01}_{-0.01}$ & $M_{\odot}$ \\
    Ejecta Mass ($M_{\mathrm{ej}}$) & $0.42^{+0.01}_{-0.01}$ & $0.49^{+0.02}_{-0.02}$ & $M_{\odot}$ \\
    $^{56}$Ni Mass ($M_{\mathrm{Ni}}$) & 1.41 $\times$ 10$^{-2}$ $\pm$ 1 $\times$ 10$^{-4}$  & 1.40 $\times$ 10$^{-2}$ $\pm$ 1 $\times$ 10$^{-4}$ & $M_{\odot}$ \\
    Explosion Epoch ($t_{\rm exp}$)    & $-10.99^{+0.01}_{-0.01}$ & $-10.98^{+0.02}_{-0.01}$  & days \\
    Shock Velocity ($v_{s}$) & $1.06^{+0.29}_{-0.25}$ & N/A  & 10$^{4}$ km s$^{-1}$\\
    \hline
    \end{tabular}
    \vspace{0.2cm}
    \caption{Summary of MCMC fit parameters with shock-cooling envelope models as described by \cite{Piro21} and \cite{Margalit22} along with power from radioactive decay of $^{56}$Ni. Explosion epoch is calculated from the second peak.}
    \label{tab:bolometric_fitting}
\end{table*}

\begin{figure*}
    \centering
    \includegraphics[width=\textwidth]{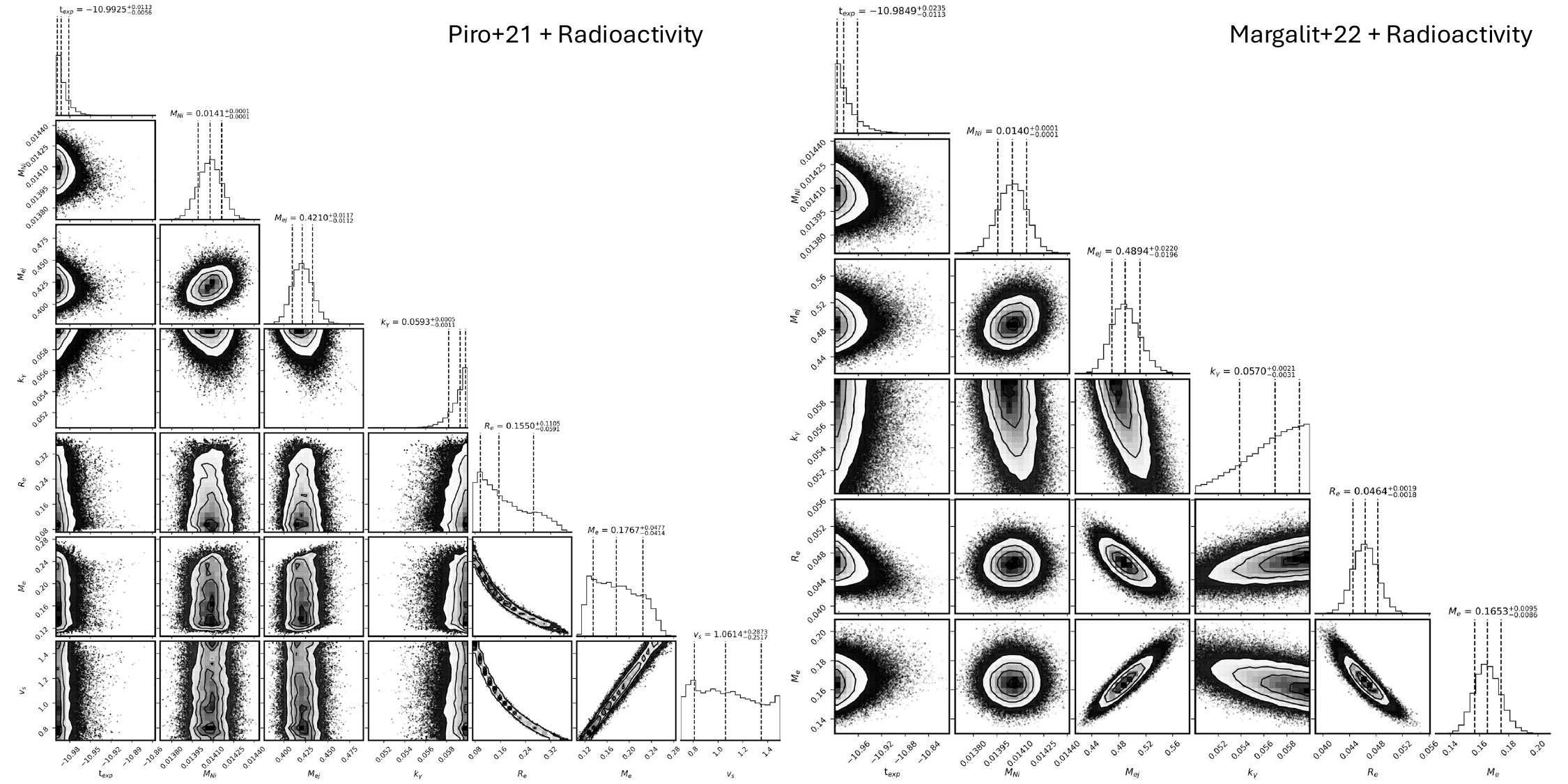}
    \caption{Corner plots showing covariance between fitted parameters in both two-component fits as described in Section \ref{sec:7.1}.}
    \label{fig:corner_plots}
\end{figure*}


\bibliography{sample7}{}
\bibliographystyle{aasjournalv7}


\end{CJK*}
\end{document}